\documentclass[twocolumn]{aastex63}
\usepackage{subfigure}
\usepackage{url}
\usepackage{hyperref}
\usepackage{epsfig}
\usepackage{graphicx}
\usepackage{sidecap}
\usepackage{hanging}
\usepackage{color}
\usepackage{multirow}
\usepackage{enumerate}
\usepackage{natbib}
\usepackage{amssymb}
\usepackage{amsmath}
\usepackage[bottom]{footmisc}

\newcommand{\jatis}{JATIS}

\newcommand{\Kepler}{{Kepler}}

\newcommand{\Gaia}{{\it Gaia}}
\newcommand{\TESS}{{TESS}}

\providecommand{\e}[1]{\ensuremath{\, \times \, 10^{#1}}}

\shorttitle{Imaging Characterization of RV Exoplanets}
\shortauthors{Dalba et al.}

\begin{document}

\title{Speckle Imaging Characterization of Radial Velocity Exoplanet Systems}

\correspondingauthor{Paul A. Dalba}
\email{pdalba@ucr.edu}


\author[0000-0002-4297-5506]{Paul A. Dalba}
\altaffiliation{NSF Astronomy \& Astrophysics Postdoctoral Fellow}
\affiliation{Department of Earth and Planetary Sciences, University of California Riverside, 900 University Avenue, Riverside, CA 92521, USA}

\author[0000-0002-7084-0529]{Stephen R. Kane}
\affiliation{Department of Earth and Planetary Sciences, University of California Riverside, 900 University Avenue, Riverside, CA 92521, USA}

\author[0000-0002-2532-2853]{Steve B. Howell}
\affiliation{NASA Ames Research Center, Moffett Field, CA 94035, USA}

\author[0000-0003-2159-1463]{Elliott P. Horch}
\affiliation{Department of Physics, Southern Connecticut State University, New Haven, CT 06515, USA}

\author[0000-0002-4860-7667]{Zhexing Li}
\affiliation{Department of Earth \& Planetary Sciences, University of California Riverside, 900 University Avenue, Riverside, CA 92521, USA}

\author[0000-0001-8058-7443]{Lea A. Hirsch}
\affiliation{Kavli Institute for Particle Astrophysics and Cosmology, Stanford University, Stanford, CA 94305, USA}

\author[0000-0002-0040-6815]{Jennifer Burt}
\affiliation{Jet Propulsion Laboratory, California Institute of Technology, 4800 Oak Grove Drive, Pasadena, CA 91109, USA}

\author[0000-0003-2630-8073]{Timothy D. Brandt}
\affiliation{Department of Physics, University of California, Santa Barbara, Santa Barbara, CA 93106, USA}

\author[0000-0003-4603-556X]{Teo Mo\v{c}nik}
\affiliation{Gemini Observatory/NSF's NOIRLab, 670 N. A'ohoku Place, Hilo, HI 96720, USA}

\author[0000-0003-4155-8513]{Gregory W. Henry}
\affiliation{Center of Excellence in Information Systems, Tennessee State University, Nashville, TN 37209, USA}

\author[0000-0002-0885-7215]{Mark E. Everett}
\affiliation{National Optical Astronomy Observatory, Tucson, AZ 85719, USA}

\author[0000-0001-8391-5182]{Lee J. Rosenthal}
\affiliation{Department of Astronomy, California Institute of Technology, Pasadena, CA 91125, USA}

\author[0000-0001-8638-0320]{Andrew W. Howard}
\affiliation{Department of Astronomy, California Institute of Technology, Pasadena, CA 91125, USA}


\begin{abstract}

We conducted speckle imaging observations of 53 stellar systems that were members of long-term radial velocity (RV) monitoring campaigns and exhibited substantial accelerations indicative of planetary or stellar companions in wide orbits. Our observations were made with blue and red filters using the Differential Speckle Survey Instrument at Gemini-South and the NN-Explore Exoplanet Stellar Speckle Imager at the WIYN telescope. The speckle imaging identifies eight luminous companions within 2$\arcsec$ of the primary stars. In three of these systems---HD~1388, HD~87359, and HD~104304---the properties of the imaged companion are consistent with the RV measurements, suggesting that these companions may be associated with the primary and the cause of the RV variation. For all 53 stellar systems, we derive differential magnitude limits (i.e., contrast curves) from the imaging. We extend this analysis to include upper limits on companion mass in systems without imaging detections. In 25 systems, we rule out companions with masses greater than 0.2~$M_{\sun}$, suggesting that the observed RV signals are caused by late-M dwarfs or substellar (potentially planetary) objects. On the other hand, the joint RV and imaging analysis almost entirely rules out planetary explanations of the RV signal for HD~19522 and suggests that the companion must have an angular separation below a few tenths of an arcsecond. This work highlights the importance of combined RV and imaging observations for characterizing the outer regions of nearby planetary systems.

\end{abstract}


 
\section{Introduction} \label{sec:intro}

Indirect detection of exoplanets has yielded thousands of new discoveries, thanks largely to the efforts of large scale surveys that have successfully monitored thousands of stars. In particular, the radial velocity (RV) and transit detection techniques have contributed the bulk of these discoveries via missions such as \Kepler\ \citep{Borucki2016}, the Transiting Exoplanet Survey Satellite \citep[\TESS;][]{Ricker2015}, and numerous ground-based surveys \citep[e.g.,][]{Christian2006,Howard2010,Hojjatpanah2019}. However, the validation of exoplanet candidates can be an expensive endeavor, often requiring follow-up observations with competitive facilities and detailed analyses of ancillary data sets \citep{Santerne2015,Parvianainen2019,Torres2017}. For RV detections of exoplanet candidates, the use of high resolution imaging can reveal the presence of stellar companions, thus resolving an inclination ambiguity to the companion mass or the nature of a long-term RV trend \citep{Crepp2012,Wittrock2016,Kane2019b}. Such imaging can occasionally reveal the presence of exotic stellar companions, whose low luminosity and RV signature can mimic that of a planet \citep{Crepp2013,Kane2019a}.

Several RV surveys have now been operating for a few decades, resulting in a sensitivity to long-period companions that are potentially Jupiter and Saturn analogs \citep{Wittenmyer2020,Feng2019,Rowan2016,Boisse2012}. Long-period companions to relatively close stars can translate into large angular separations, enabling the potential for direct detection if the companions are stellar \citep{Cheetham2018}. Even companions in eccentric orbits can be detected since the fraction of the orbit outside of the inner working angle and close to apastron can comprise a substantial fraction of the full orbital period \citep{Kane2013}. Thus, the results of RV surveys for exoplanets published by \citet{Butler2017} present an opportunity to validate numerous exoplanet candidates through a corresponding imaging survey of the host stars. Fortunately, many of these candidates have also been monitored by \citet{Rosenthal2021}, providing further constraints on the companion orbits and mass estimates.

Here we present the results of a speckle imaging survey for 53 stars that have proposed substellar companions based on the RV data of \citet{Butler2017} and \citet{Rosenthal2021}. Our imaging data are used to constrain the masses of the detected RV companions and search for possible stellar companions. In Section~\ref{sec:targ} we describe our target sample. In Section~\ref{sec:speckle}, we summarize the speckle imaging observations that were conducted for those stars. Section~4 discusses the utilization of RV orbital solutions in combination with our speckle imaging data. The resulting mass constraints for the detected companions are provided in Section~\ref{sec:results}. We discuss the implications of our results for current and future exoplanets surveys in Section~\ref{sec:discussion} and provide a summary and concluding remarks in Section~\ref{sec:conclusions}.


\section{Target Sample}\label{sec:targ}

Our target list consists of a subset of the systems identified as having planet candidates by \citet{Butler2017} that are also amenable to speckle imaging. We excluded systems that showed significant correlation between the RVs and the stellar activity as determined by the model comparison technique of \citet{Butler2017}. Such targets were identified in Table 2 of \citet{Butler2017} as having ``Activity'' as their interpretation. We only chose to observe targets with the interpretation of ``Candidate,'' meaning that the planet candidate model was favored over a stellar activity model to 0.1\% false alarm probability. 

During the preparation of this paper, many of these systems were further characterized through the California Planet Search (CPS) Legacy Survey, which published decades of RV observations from the High Resolution Echelle Spectrometer (HIRES) at the Keck~I telescope \citep{Rosenthal2021}. We have included those results in the following analysis of imaging data.

The stellar properties needed for our analysis included $V$-band apparent magnitude, distance, and mass (see Table \ref{tab:stellar}). We collected $V$-band magnitudes for each star from the online Simbad database\footnote{\url{http://simbad.u-strasbg.fr/simbad/}.}. The distance to each star was determined using \Gaia\ Data Release 2 (DR2) parallax measurements \citep{Gaia2018} including a prior on distance \citet{BailerJones2018}. We adopted the stellar masses of \citet{Rosenthal2021} for those stars that are members of the CPS legacy survey. For the remaining stars, we processed archival HIRES spectra following \citet{Fulton2015b} to infer the stellar mass.

\startlongtable
\begin{deluxetable}{lcccc}
    \tabletypesize{\scriptsize}
    \tablecaption{Summary of Stellar Properties for Target Sample} 
    \tablecolumns{5}
    \tablewidth{0pt}
    \tablehead{  
        \colhead{Star} &
        \colhead{$V$} & 
        \colhead{$d$ (pc)} &
        \colhead{$M_{\star}$ ($M_{\sun}$)} & 
        \colhead{Ref.}}
    \startdata 
    GL~317 & $11.9$ & $15.197\pm0.013$ & $0.453\pm0.009$ & 1 \\
    HD~1326 & $8.1$ & $3.5623\pm0.0006$ & $0.400\pm0.008$ & 1 \\
    HD~1388 & $6.51$ & $26.923\pm0.038$ & $1.027\pm0.046$ & 1 \\
    HD~1461 & $6.47$ & $23.453\pm0.031$ & $1.031\pm0.047$ & 1 \\
    HD~3765 & $7.36$ & $17.926\pm0.032$ & $0.852\pm0.033$ & 1 \\
    HD~5319 & $8.05$ & $121.41\pm0.7$ & $1.53\pm0.14$ & 2 \\
    HD~6558 & $8.2$ & $81.89\pm0.42$ & $1.29\pm0.033$ & 1 \\
    HD~6734 & $6.44$ & $46.73\pm0.11$ & $0.968\pm0.091$ & 1 \\
    HD~7924 & $7.17$ & $16.9922\pm0.0072$ & $0.802\pm0.033$ & 1 \\
    HD~9986 & $6.77$ & $25.445\pm0.026$ & $1.032\pm0.05$ & 1 \\
    HD~10436 & $7.75$ & $13.5098\pm0.0065$ & $0.632\pm0.015$ & 1 \\
    HD~16160 & $5.8$ & $7.2339\pm0.0076$ & $0.752\pm0.026$ & 1 \\
    HD~19522 & $8.11$ & $102.16\pm0.81$ & $1.28\pm0.03$ & 2 \\
    HD~24040 & $7.5$ & $46.62\pm0.14$ & $1.104\pm0.053$ & 1 \\
    HD~25311 & $8.28$ & $105.96\pm0.56$ & $1.4\pm0.04$ & 2 \\
    HD~34445 & $7.31$ & $46.09\pm0.1$ & $1.11\pm0.06$ & 1 \\
    HD~42618 & $6.85$ & $24.336\pm0.025$ & $0.92\pm0.046$ & 1 \\
    HD~50499 & $7.21$ & $46.285\pm0.056$ & $1.253\pm0.035$ & 1 \\
    HD~55696 & $7.95$ & $77.97\pm0.18$ & $1.36\pm0.03$ & 2 \\
    HD~68017 & $6.78$ & $21.573\pm0.027$ & $0.815\pm0.014$ & 1 \\
    HD~68988 & $8.2$ & $60.84\pm0.19$ & $1.172\pm0.049$ & 1 \\
    HD~72490 & $7.82$ & $126.3\pm1.1$ & $1.37\pm0.15$ & 2 \\
    HD~75732 & $5.96$ & $12.586\pm0.012$ & $0.975\pm0.045$ & 1 \\
    HD~75898 & $8.03$ & $78.05\pm0.3$ & $1.29\pm0.06$ & 2 \\
    HD~83443 & $8.23$ & $40.899\pm0.063$ & $1.007\pm0.045$ & 1 \\
    HD~87359 & $7.49$ & $31.268\pm0.046$ & $0.982\pm0.05$ & 1 \\
    HD~92788 & $7.31$ & $34.654\pm0.06$ & $1.076\pm0.044$ & 1 \\
    HD~94834 & $7.6$ & $98.16\pm0.64$ & $1.39\pm0.15$ & 2 \\
    HD~95735 & $7.5$ & $2.5484\pm0.0059$ & $0.392\pm0.008$ & 1 \\
    HD~99491 & $6.49$ & $18.199\pm0.015$ & $1.02\pm0.044$ & 1 \\
    HD~104304 & $5.54$ & $12.693\pm0.02$ & $1.026\pm0.045$ & 1 \\
    HD~111031 & $6.87$ & $31.206\pm0.051$ & $1.099\pm0.046$ & 1 \\
    HD~114174 & $6.78$ & $26.355\pm0.036$ & $0.968\pm0.044$ & 1 \\
    HD~114783 & $7.56$ & $21.063\pm0.028$ & $0.867\pm0.036$ & 1 \\
    HD~126614 & $8.81$ & $73.1\pm0.25$ & $1.021\pm0.033$ & 1 \\
    HD~129814 & $7.52$ & $41.95\pm0.11$ & $0.973\pm0.043$ & 1 \\
    HD~145675 & $6.61$ & $17.9323\pm0.0073$ & $0.969\pm0.042$ & 1 \\
    HD~146233 & $5.49$ & $14.125\pm0.023$ & $0.995\pm0.044$ & 1 \\
    HD~156668 & $8.42$ & $24.332\pm0.017$ & $0.785\pm0.024$ & 1 \\
    HD~180053 & $7.93$ & $137.04\pm0.61$ & $2.02\pm0.05$ & 2 \\
    HD~188015 & $8.24$ & $50.67\pm0.11$ & $1.043\pm0.048$ & 1 \\
    HD~190406 & $5.8$ & $17.713\pm0.022$ & $1.07\pm0.044$ & 1 \\
    HD~195564 & $5.8$ & $24.746\pm0.057$ & $1.121\pm0.034$ & 1 \\
    HD~197076 & $6.43$ & $20.886\pm0.016$ & $0.979\pm0.05$ & 1 \\
    HD~197162 & $8.01$ & $141.21\pm0.68$ & $1.2\pm0.17$ & 2 \\
    HD~202696 & $8.23$ & $188.5\pm1.6$ & $1.86\pm0.24$ & 2 \\
    HD~207077 & $8.24$ & $155.5\pm1.6$ & $1.35\pm0.14$ & 2 \\
    HD~216520 & $7.53$ & $19.552\pm0.011$ & $0.791\pm0.03$ & 1 \\
    HD~217850 & $8.52$ & $65.8\pm0.87$ & $1.09\pm0.05$ & 2 \\
    HD~221354 & $6.76$ & $16.8686\pm0.0089$ & $0.864\pm0.027$ & 1 \\
    HD~265866 & $10.1$ & $5.5806\pm0.002$ & $0.373\pm0.009$ & 1 \\
    HIP~52942~A & $9.29$ & $164.1\pm1.3$ & $1.223\pm0.072$ & 1 \\
    HIP~57050 & $12.0$ & $11.0143\pm0.0064$ & $0.374\pm0.009$ & 1 \\
    \enddata
    \tablenotetext{}{Notes. All distances were adopted from \citet{BailerJones2018}. In the references column, 1 refers to masses adopted from \citet{Rosenthal2021}, and 2 refers to masses inferred by modeling archival HIRES spectra following \citet{Fulton2015b}.}
    \label{tab:stellar}
\end{deluxetable}


\section{Speckle Imaging Observations}\label{sec:speckle}

We summarize the speckle imaging observations in Table \ref{tab:imaging}. Observations were acquired using the Differential Speckle Survey Instrument \citep[DSSI,][]{Horch2009,Horch2011} at the Gemini South Telescope and the NN-Explore Exoplanet Stellar Speckle Imager \citep[NESSI,][]{Scott2018} at the Wisconsin-Indiana-Yale-NOAO (WIYN) Telescope.

\startlongtable
\begin{deluxetable}{lcc}
    \tablecaption{Summary of the Imaging Observations \label{tab:imaging}} 
    \tabletypesize{\scriptsize}
    \tablecolumns{3}
    \tablewidth{0pt}
    \tablehead{  
        \colhead{Star} &
        \colhead{Tel./Inst.} &  
        \colhead{Date (UT)}   
        }
    \startdata
    GL~317      & Gemini-S/DSSI & 2018 Mar 31  \\
    GL~1326     & Gemini-S/DSSI & 2018 Feb 01\\
    HD~1388     & WIYN/NESSI    & 2017 Sep 03  \\
    HD~1461     & WIYN/NESSI    & 2017 Sep 03  \\
    HD~3765     & WIYN/NESSI    & 2017 Sep 03  \\
    HD~5319     & WIYN/NESSI    & 2017 Sep 03  \\
    HD~6558     & WIYN/NESSI    & 2017 Sep 04  \\
    HD~6734     & WIYN/NESSI    & 2017 Sep 04  \\
    HD~7924     & WIYN/NESSI    & 2017 Sep 03  \\
    HD~9986     & WIYN/NESSI    & 2017 Sep 04  \\
    HD~10436    & WIYN/NESSI    & 2017 Sep 03  \\
    HD~16160    & WIYN/NESSI    & 2018 Feb 02  \\
    HD~19522    & WIYN/NESSI    & 2018 Feb 02  \\
    HD~24040    & WIYN/NESSI    & 2017 Sep 03  \\
    HD~25311    & WIYN/NESSI    & 2018 Feb 06  \\
    HD~34445    & WIYN/NESSI    & 2017 Apr 05  \\
    HD~42618    & WIYN/NESSI    & 2017 Apr 05  \\
    HD~50499    & Gemini-S/DSSI & 2018 Mar 30  \\
    HD~55696    & Gemini-S/DSSI & 2018 Mar 30  \\
    HD~68017    & WIYN/NESSI    & 2017 Mar 13  \\
    HD~68988    & WIYN/NESSI    & 2017 Mar 13   \\
    HD~72490    & WIYN/NESSI    & 2017 Mar 13  \\
    HD~75732    & WIYN/NESSI    & 2018 Feb 03  \\
    HD~75898    & WIYN/NESSI    & 2017 Mar 13  \\
    HD~83443    & Gemini-S/DSSI & 2018 Mar 31   \\
    HD~87359    & WIYN/NESSI    & 2017 Apr 09 \\
    HD~92788    & WIYN/NESSI    & 2017 Apr 09  \\
    HD~94834    & WIYN/NESSI    & 2017 Apr 05  \\
    HD~95735    & WIYN/NESSI    & 2018 Feb 02  \\
    HD~99491    & WIYN/NESSI    & 2017 Apr 05 \\
    HD~104304   & WIYN/NESSI    & 2018 Feb 02 \\
    HD~111031   & WIYN/NESSI    & 2018 Feb 01 \\
    HD~114174   & WIYN/NESSI    & 2017 May 22 \\
    HD~114783   & WIYN/NESSI    & 2017 May 22   \\
    \multirow{2}{*}{HD~126614}  & WIYN/NESSI    & 2018 Jun 19  \\
                & Gemini-S/DSSI & 2018 Mar 31 \\
    HD~129814   & WIYN/NESSI    & 2017 Mar 13 \\
    HD~145675   & WIYN/NESSI    & 2017 Aug 09 \\
    HD~146233   & WIYN/NESSI    & 2017 Aug 14 \\
    HD~156668   & WIYN/NESSI    & 2017 Aug 12  \\
    HD~180053   & WIYN/NESSI    & 2017 Sep 05  \\
    HD~188015   & WIYN/NESSI    & 2017 Aug 09 \\
    HD~190406   & WIYN/NESSI    & 2017 Aug 12 \\
    HD~195564   & WIYN/NESSI    & 2017 Sep 04 \\
    HD~197076   & WIYN/NESSI    & 2017 Aug 12 \\
    HD~197162   & WIYN/NESSI    & 2017 Sep 02 \\
    HD~202696   & WIYN/NESSI    & 2017 Aug 12 \\    
    HD~207077   & WIYN/NESSI    & 2017 Aug 12 \\
    HD~216520   & WIYN/NESSI    & 2017 Sep 03  \\
    HD~217850   & WIYN/NESSI    & 2017 Aug 12  \\
    HD~221354   & WIYN/NESSI    & 2017 Aug 12  \\
    HD~265866   & WIYN/NESSI    & 2017 Apr 05  \\
    HIP~52942~A & WIYN/NESSI    & 2017 Apr 05 \\ 
    HIP~57050   & WIYN/NESSI    & 2017 Apr 09   \\
    \enddata
\end{deluxetable} 

Standard procedure was followed when acquiring speckle imaging observations with either instrument \citep{Horch2011,Howell2011}. In both cases, the image at the telescope focal plane is collimated and then separated into red and blue components using a dichroic beamsplitter. The light is then collected by two electron-multiplying charge-coupled devices (EMCCDs) using sequences of short (60~ms at Gemini-South, 40~ms at WIYN) exposures. For DSSI, the central wavelength and bandwidth for the blue and red filters are 692 and 40 nm, and 880 and 50 nm, respectively. For NESSI, the central wavelength and bandwidth for the blue and red filters are 562 and 44 nm, and 832 and 40 nm, respectively. 

Full descriptions of the reduction and analysis techniques applied to the DSSI and NESSI data are provided by \citet{Horch2011} and \citet{Howell2011}, respectively. The reduction pipeline for NESSI is based on the one developed for DSSI when it began to be used for exoplanet follow-up observations at WIYN and Gemini observatories. In both cases, reconstructed images with angular resolution near the diffraction limit are created to enable the identification of stellar companions at or beyond $\sim 0\farcs1$ from the primary star. These companions are found by examining the statistics of local maxima and minima in the image as a function of angular separation from the primary \citep{Horch2011}. In concentric annuli 0$\farcs$1 in radius surrounding the primary, we estimate the $5\sigma$ detection limit in terms of instrumental magnitude difference ($\Delta m_i$) between the primary and a possible companion, where $i$ is the filter. We assume that $\Delta m_i$ approaches zero at the diffraction limit. Since we are attempting to discover new companion stars \citep[as opposed to detecting those already known to exist;][]{Horch2019}, this assumption ensures that any companions will produce their own peaks in the image. Any peak that exceeds the $5\sigma$ value of $\Delta m_i$ at a specific angular distance is examined as a possible stellar companion. We consider peaks at similar angular separations in both filters as strong evidence for a stellar companion. However, M dwarf companions are occasionally only detected in the red filter. In Table \ref{tab:det_limits}, we list the $5\sigma$ detection limit for each of our target stars at $0\farcs1$ and $1\farcs0$. 

\begin{figure*}
  \centering
    \begin{tabular}{ccc}
      \includegraphics[width=5.6cm]{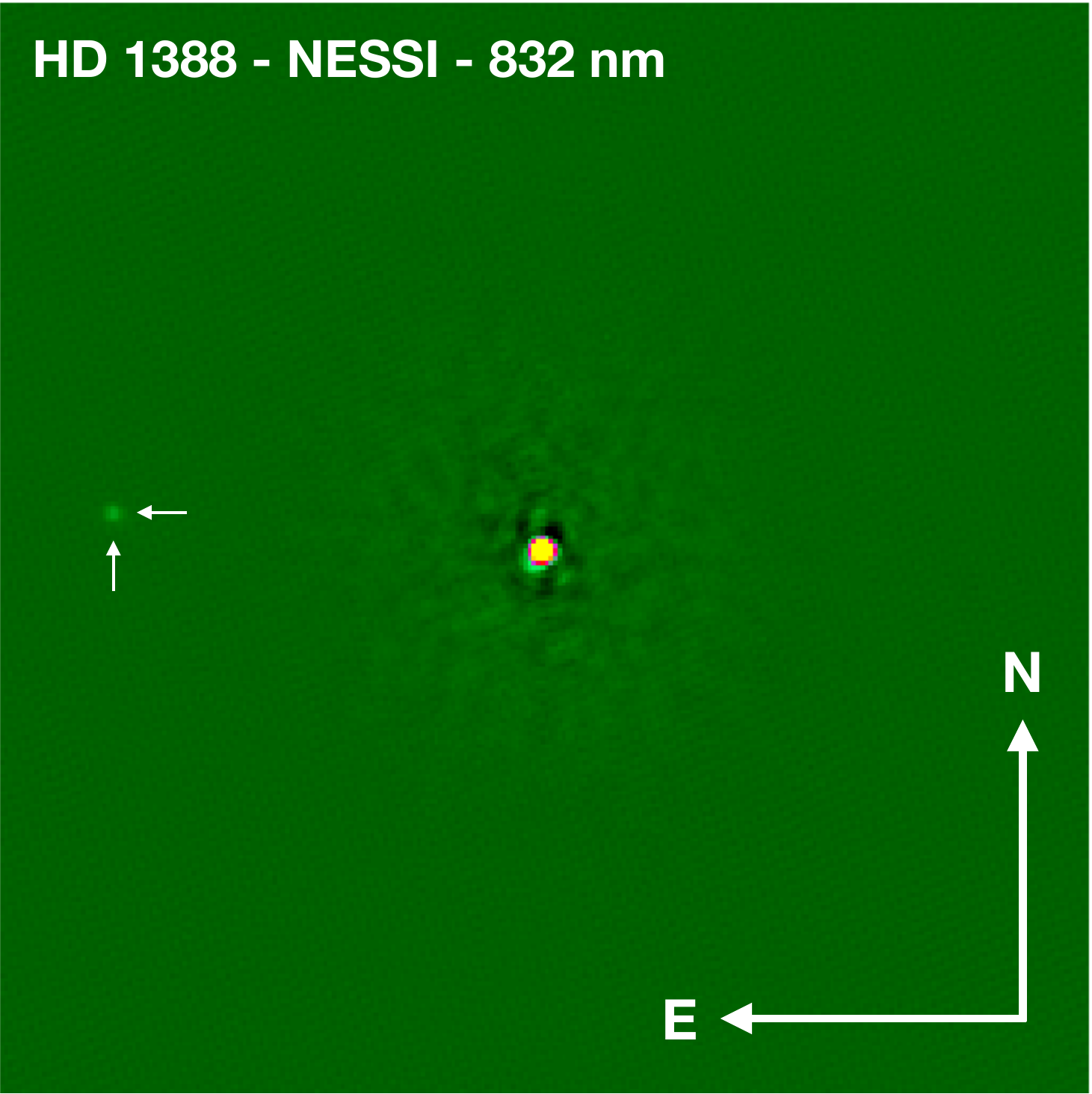} &
      \includegraphics[width=5.6cm]{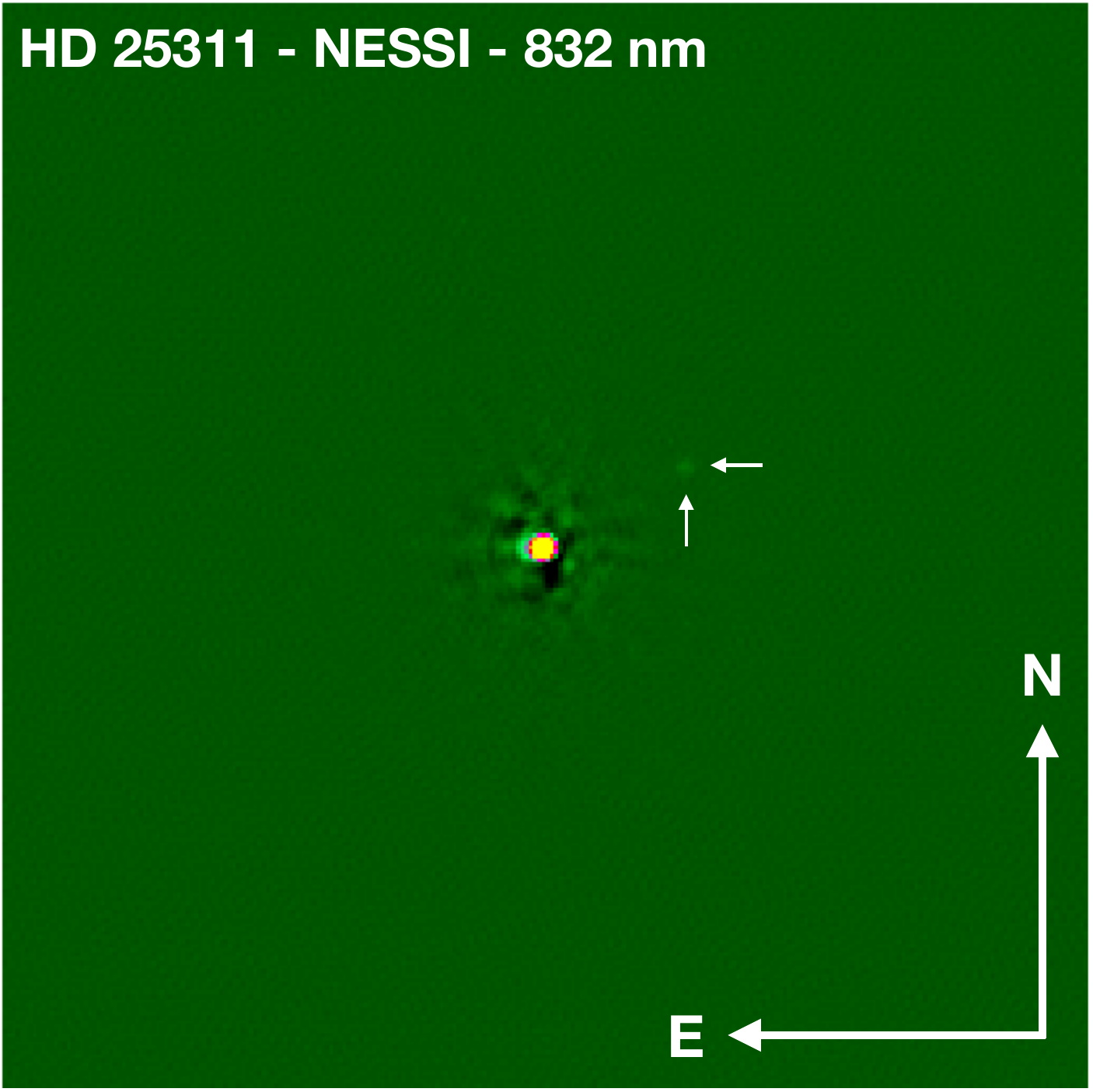} &
      \includegraphics[width=5.6cm]{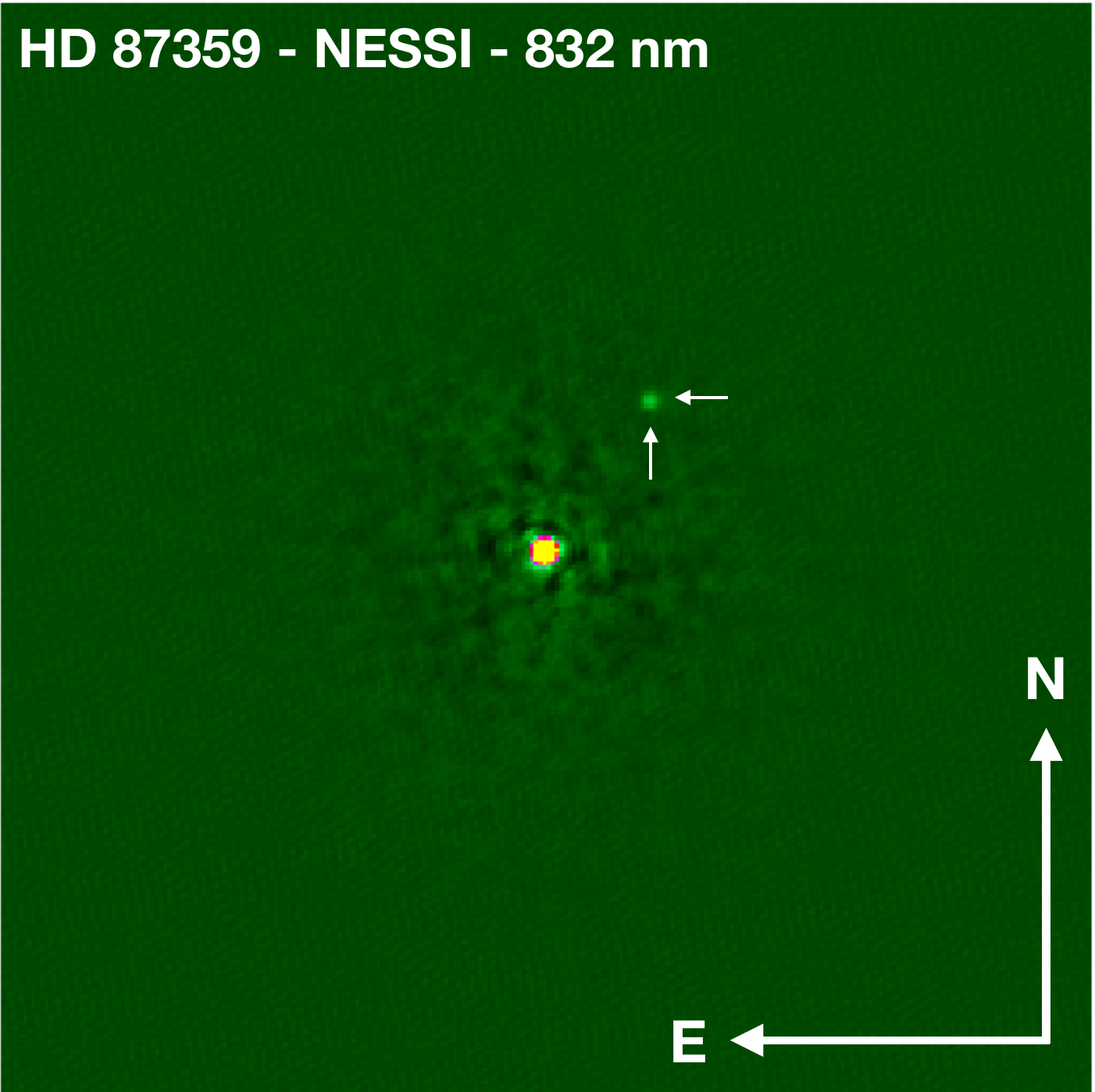} \\
      \includegraphics[width=5.6cm]{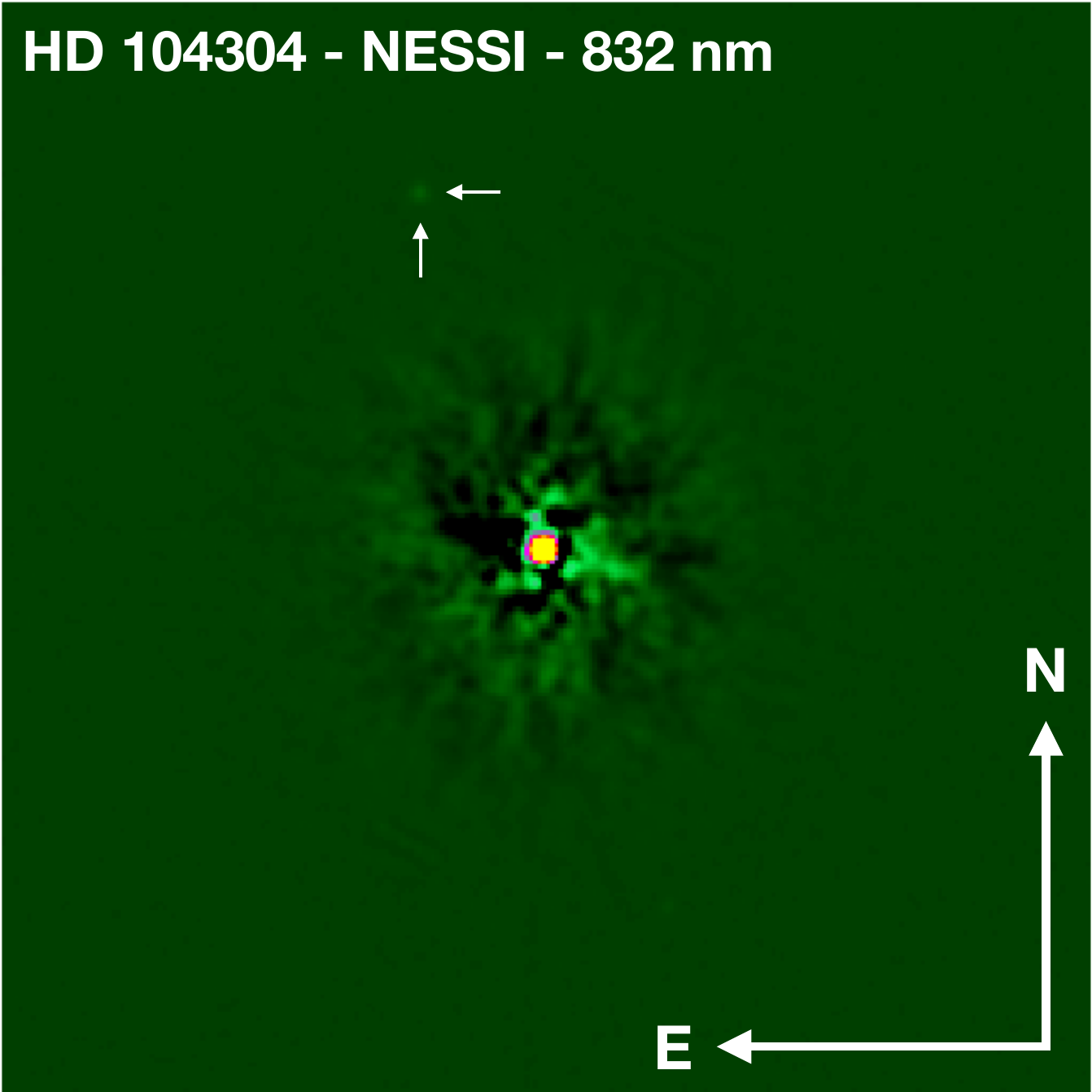} &
      \includegraphics[width=5.6cm]{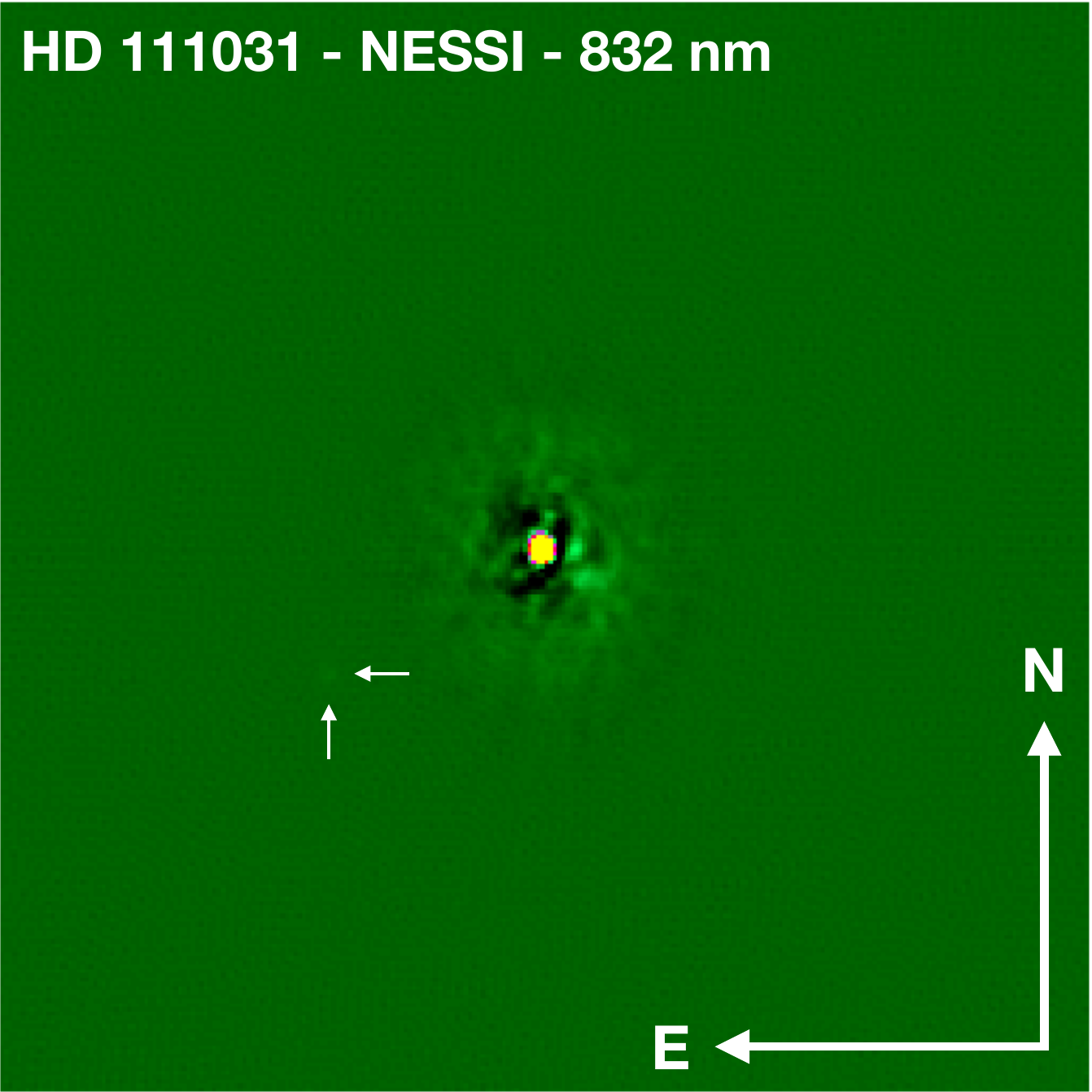} &
      \includegraphics[width=5.6cm]{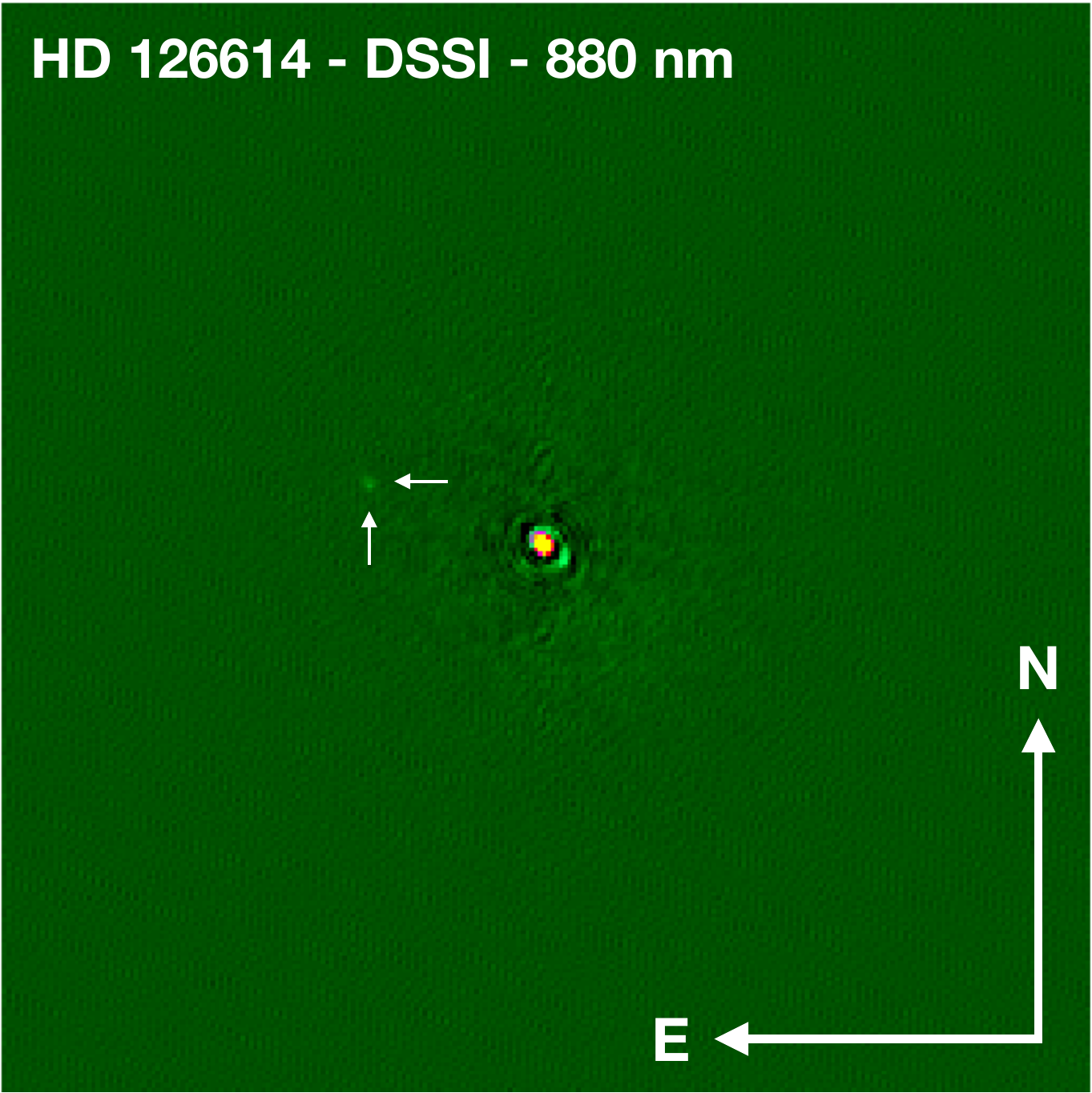} \\
      \includegraphics[width=5.6cm]{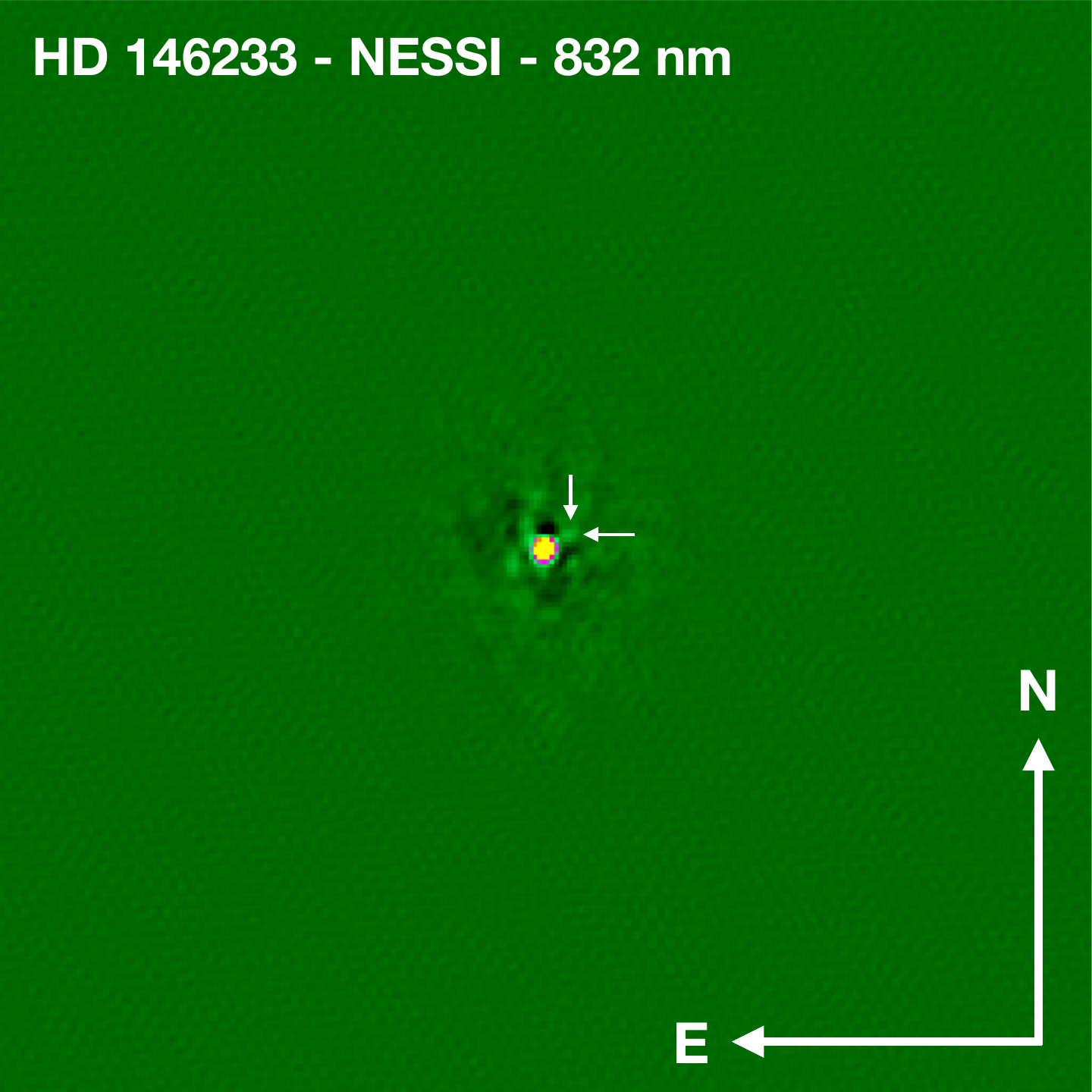} &
      \includegraphics[width=5.6cm]{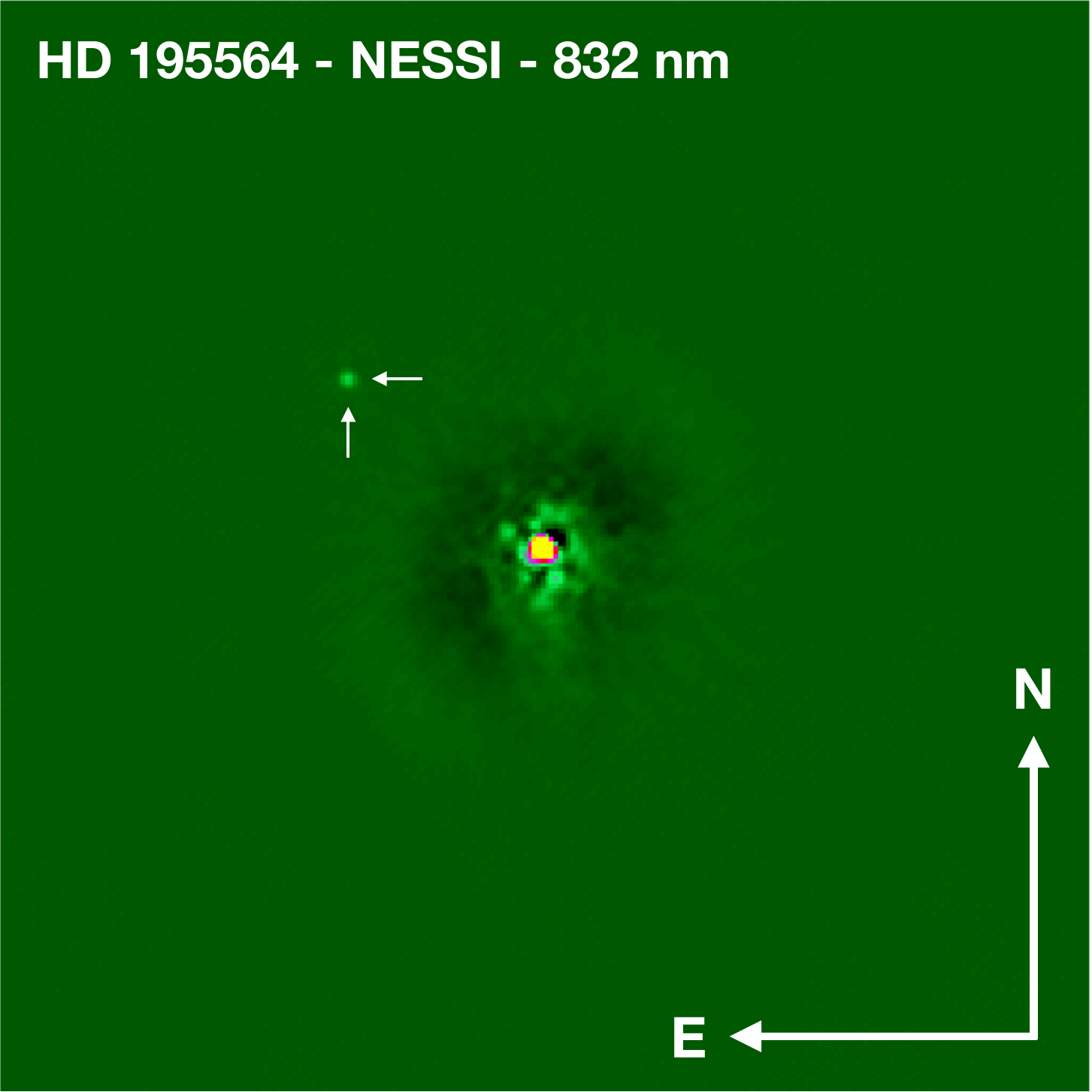} &
    \end{tabular}
  \caption{Speckle images for systems with detected companions (indicated by the small arrows). The field of view for each image is 4$\farcs$6 on each side.}
  \label{fig:companions}
\end{figure*}

We identified luminous companions in the speckle imaging observations of eight stars in the sample. The processed speckle images for these eight systems are shown in Figure \ref{fig:companions}. The companions are readily detectable by eye for each system. Various observed properties of the companions including the position angle (PA), angular separation ($\alpha$), and the product of seeing and separation \citep[$q^{\prime}$;][]{Horch2004,Horch2011} are listed in Table \ref{tab:comp_obs}. For PA and $\alpha$, we adopt representative uncertainties from \citet{Horch2019}. In Table \ref{tab:comp_obs}, we also list the proper motion of the primary star in RA and Dec ($\mu_{\alpha}$ and $\mu_{\delta}$, respectively) as measured by \Gaia\ \citep{Gaia2018}.
We recover two known binary systems: HD~126614 \citep{Howard2010} and HD~195564, which is a member of the Washington Double Star Catalog with observations dating back to 1878 \citep[e.g.,][]{Burnham1879,Lloyd2002,Tokovinin2016}. We also note that the speckle imaging was not sensitive enough to detect the known white dwarf companion in the HD~114174 system \citep{Crepp2013}. Lastly, one star in our sample, HIP~52942~A, is a known wide binary star, with a common proper motion companion almost due west at a separation of 17$\farcs$6 \citep{Gaia2018}. This companion is well beyond the field of view of our speckle imaging. Instead, our data searched for a tertiary star orbiting the A component of the binary.

\begin{deluxetable*}{lcccccccccc}
    \tablecaption{Observed Properties of the Imaged Companions\label{tab:comp_obs}} 
    \tablecolumns{9}
    \tablewidth{0pt}
    \tablehead{  
        \colhead{Star} &
        \colhead{$\mu_{\alpha}$ (mas~yr$^{-1}$)} &
        \colhead{$\mu_{\delta}$ (mas~yr$^{-1}$)} &
        \colhead{Instrument} & 
        \colhead{Filter (nm)} & 
        \colhead{Seeing ($\arcsec$)} &
        \colhead{PA ($\degr$)} &
        \colhead{$\alpha$ ($\arcsec$)} &
        \colhead{$\Delta m_i$} &
        \colhead{$q^{\prime}$ (arcsec$^2$)} 
        }
    \startdata
    HD~1388 & $401.216$ & $-0.129$ & NESSI & 832 & 1.19 & 85.8$\pm$0.6 & 1.8459$\pm0.0013$ & 4.98 & 2.197  \\
    HD~25311 & $-4.709$ & $-90.048$ & NESSI & 832 & 0.36 & 300.9$\pm$0.6 & 0.6954$\pm0.0013$ & 6.16 & 0.250  \\
    HD~87359 & $132.648$ & $-207.134$ & NESSI & 832 & 1.18 & 324.9$\pm$0.6 & 0.7818$\pm0.0013$ & 4.48 & 0.923     \\
    HD~104304 & $82.841$ & $-482.807$ & NESSI & 832 & 0.59 & 18.9$\pm$0.6 & 1.6238$\pm0.0013$ & 6.65 & 0.958  \\
    HD~111031 & $-279.792$ & $46.712$ & NESSI & 832 & 0.45 & 121.3$\pm$0.6 & 1.0558$\pm0.0013$ & 6.91 & 0.475  \\
    \multirow{2}{*}{HD~126614} & \multirow{2}{*}{$-149.881$} & \multirow{2}{*}{$-145.915$} & DSSI & 880 & 0.58 & 75.7$\pm$0.6 & 0.4830$\pm0.0013$ & 4.82 & 0.281  \\
    &&& NESSI & 832 & 0.65 & 76.7$\pm$0.6 & 0.4920$\pm0.0013$ & 4.93 & 0.318    \\
    \multirow{2}{*}{HD~146233} & \multirow{2}{*}{$-11.305$} & \multirow{2}{*}{$-9.062$} & NESSI & 562 & 0.72 & 293.9$\pm$0.6 & 0.1336$\pm0.0013$ & 3.57 & 0.096  \\
     &&& NESSI & 832 & 0.58 & 296.1$\pm$0.6 & 0.1334$\pm0.0013$ & 3.59 & 0.077   \\
     \multirow{2}{*}{HD~195564} & \multirow{2}{*}{$309.542$} & \multirow{2}{*}{$110.075$} & NESSI & 562 & 0.64 & 49.4$\pm$0.6 & 1.1130$\pm0.0013$ & 6.20 & 0.712    \\
     &&& NESSI & 832 & 0.47 & 49.8$\pm$0.6 & 1.1049$\pm0.0013$ & 4.67 & 0.519     
    \enddata
    \tablenotetext{}{Notes. Proper motions refer to the primary stars. Errors on PA and $\alpha$ are representative based on \citet{Horch2019}. For systems with $q^{\prime} \gtrsim 0.6$~arcsec$^2$, $\Delta m_i$ represents an upper limit rather than a precise value. See the text for an explanation.}
\end{deluxetable*} 

We also derive several properties of the imaged companions (Table \ref{tab:comp_derived}). Once again employing a spectral library \citep{Pickles1998}, the NESSI bandpasses, and the known $V$-band magnitudes of the primaries, we estimate the spectral types of the companions based on their red and blue NESSI magnitudes. We also estimate the $V$-band magnitude differences ($\Delta m_V$) of the companions. 

The magnitude and spectral type estimates in Tables \ref{tab:comp_obs} and \ref{tab:comp_derived} depend on the photometric accuracy of the speckle imaging, which is known to degrade as a function of angular separation from the primary star \citep{Horch2011}. We quantify this with the $q^{\prime}$ parameter, which is the product of the seeing and the angular separation \citep{Horch2004}. For systems with $q^{\prime} \gtrsim 0.6$, \citet{Horch2004} and \citet{Horch2011} have shown that the derived magnitude difference is anomalously large (i.e., the companion is measured to be fainter than it actually is). In these cases, the derived $\Delta m_i$ values represent upper limits rather than precise values. Even if the magnitude of a companion is relatively uncertain, its existence and angular separation from the primary star are still valuable pieces of information when combined with the RV observations. 

\begin{deluxetable}{lcccc}
    \tablecaption{Derived Properties of the Imaged Companions\label{tab:comp_derived}} 
    \tablecolumns{5}
    \tablewidth{0pt}
    \tablehead{  
        \colhead{Star} &
        \colhead{Spectral Type} &
        \colhead{$\Delta m_V$} & 
        \colhead{Color Offset ($\sigma$)} 
        }
    \startdata
    HD~1388   & M5       & 7.8      & \nodata    \\
    HD~25311  & M6$+$    & 9.0$+$   & \nodata    \\
    HD~87359  & M5       & 7.1      & \nodata    \\
    HD~104304 & M6$+$    & 7.5$+$   & \nodata    \\
    HD~111031 & M6$+$    & 7.9$+$   & \nodata    \\
    HD~126614 & M6$+$    & 7.5$+$   & \nodata    \\
    HD~146233 & K7--M3   & 3.5--5.7 & 5.51    \\
    HD~195564 & M4--M5   & 6.5--7.5 & 0.44    
    \enddata
    \tablenotetext{}{Notes. A spectral type with a plus sign represents a lower (early-type) limit rather than a single spectral type (i.e., the companion could be a later-type star than listed here). Similarly, ranges in spectral types and $\Delta m_V$ values are provided when the data from the blue and red filters produce inconsistent results.}
\end{deluxetable}

The imaged companions in two of the eight systems are detected in both blue and red filters. This color information can be used to check for differences in the modeled and observed properties of the companion via isochrone analysis \citep{Hirsch2017}. Briefly, we use the stellar properties of the primary star (Table \ref{tab:stellar}) to determine its position on isochrones from the Dartmouth Stellar Evolution Database \citep{Dotter2008}. We then use the observed $\Delta m_i$ of the neighbor (Table \ref{tab:comp_obs}) to interpolate down the isochrone to the contrast of a hypothetical bound companion. Each filter provides its own model companion color, which we combine through a weighted average. Finally, comparing the averaged model color to the measured color, we conclude that any offset $\le3\sigma$ suggests that the off-axis source is likely associated with the primary star. 

\begin{figure*}
  \begin{center}
    \begin{tabular}{cc}
      \includegraphics[width=0.85\columnwidth]{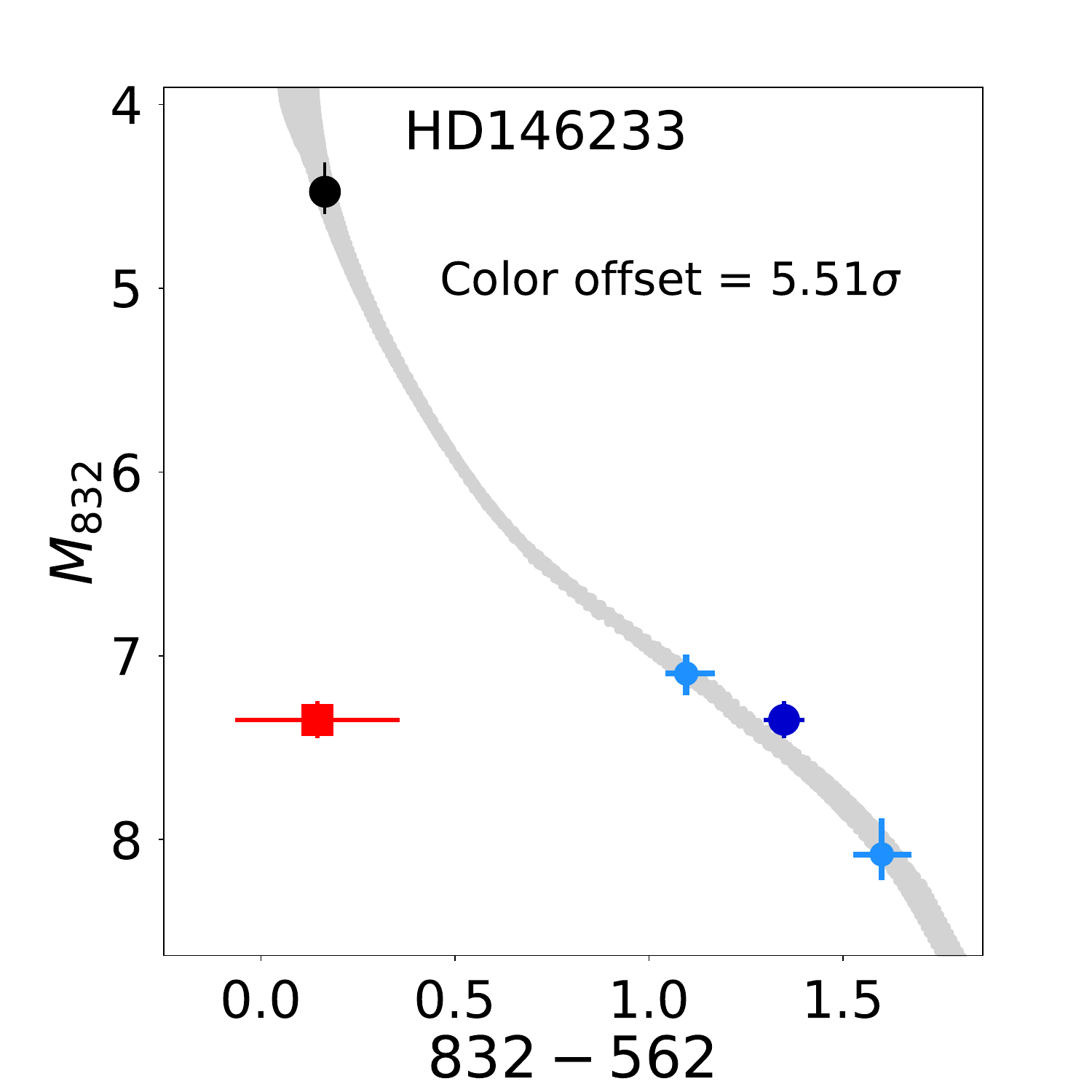} 
      \includegraphics[width=0.85\columnwidth]{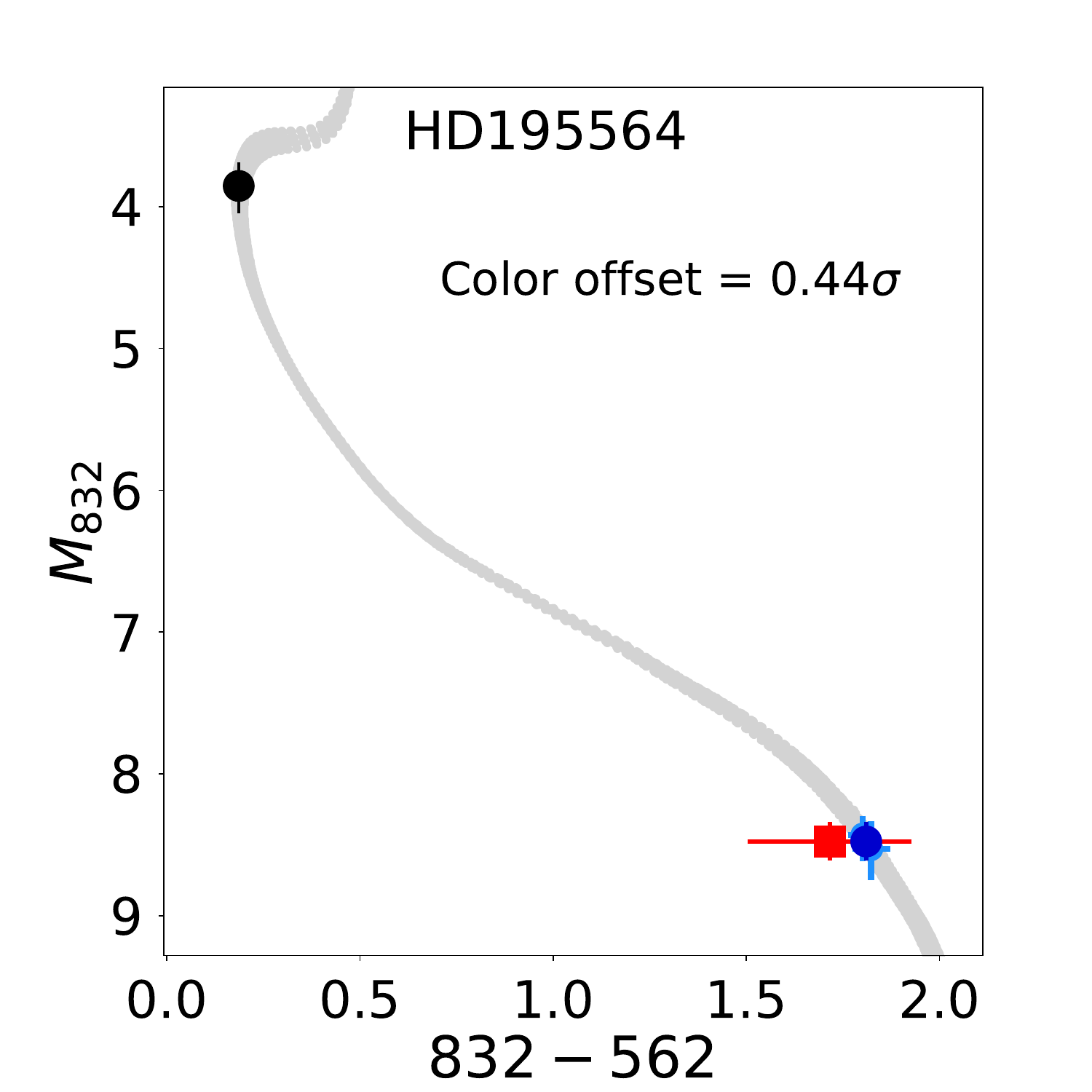} 
    \end{tabular}
  \end{center}
  \caption{Results of the isochrone analysis \citep{Hirsch2017} for the imaged companions of HD~146233 (left) and HD~195564 (right). The x-axis is the magnitude difference. The light blue points are the color predictions of the secondary based on the individual measurements of $\Delta m$ with the primary. The dark blue point is the weighted mean of the light blue points. The substantial offset between the measured (red) and modeled (dark blue) colors of the companion in the HD~146233 system suggests that it is not bound, while the opposite is true for HD~195564,} although see the text for a possible caveat for the HD~146233 system.
  \label{fig:isochrone}
\end{figure*}

The results of this isochrone analysis for HD~146233 and HD~195564 are shown in Figure \ref{fig:isochrone}. The ``color offset'' for the imaged companion of HD~195564 is small and suggests that the companion is likely bound. On the other hand, the imaged companion of HD~146233 is likely not gravitationally bound.

A detection of a companion in HD~146233 was made in each filter. However, there is doubt in the hypothesis that the detection in HD~146233 is a background star. HD~146233 has high proper motion \citep[232~mas~yr$^{1}$ in RA, $-$495~mas~yr$^{-1}$ in Dec;][]{Gaia2018} such that stellar catalogs should have detected a background star with a $V$-band magnitude of $\sim$9. However, at the location of HD~146233 at the epoch of imaging, the HST Guide Star Catalog \citep{Lasker1990,Morrison2001} and the USNO-B1.0 Catalog \citep{Monet2003} both yield nondetections within 10$\arcsec$. 

It is important to emphasize that it remains possible that our tentative detection for HD~146233 could be spurious. In that case, a possible explanation for our detection is that the limited sample statistics at small angular separations made it difficult to estimate the variance of the residuals and biased that variance estimate low \citep[e.g.,][]{Mawet2014}. A noise feature could then have been mistaken as a neighboring star. Further observations of HD~146233 are needed to confirm this result.

In addition to identifying companions, we use the $5\sigma$ magnitude limit curves to place upper limits on the mass of possible companions following \citet{Kane2014,Kane2019b}. First, the mass-luminosity relations of \citet{Henry1993} are combined with the known distance to each system to estimate the apparent V-band magnitude of a possible stellar companion as a function of mass. The apparent V-band magnitude of the primary is known (Table \ref{tab:stellar}), so we calculate the magnitude difference limits ($\Delta m_V$) of a possible companion, also as a function of its mass. We then transform $\Delta m_V$ into magnitude difference limits for blue ($\Delta m_{\rm 692}$ for DSSI and $\Delta m_{\rm 562}$ for NESSI) and red ($\Delta m_{\rm 880}$ for DSSI and $\Delta m_{\rm 832}$ for NESSI) filters using the Pickles spectral library \citep{Pickles1998} and the transmission curves of each filter. Lastly, the modeled magnitude difference limits are compared to those observed to yield upper mass limits on a companion as a function of angular separation. In Table \ref{tab:det_limits}, we list the companion mass limits that correspond to the 5$\sigma$ instrumental magnitude limit in the blue and red filters. In Figures \ref{fig:imaging1}--\ref{fig:imaging2}, we plot these limits in both filters for a subset of our targets.

\startlongtable
\begin{deluxetable*}{lcccccccc}
    \tablecaption{Summary of 5$\sigma$ Detection Limits in Magnitude from Speckle Imaging \label{tab:det_limits}} 
    \tablecolumns{9}
    \tablewidth{0pt}
    \tablehead{  
        \colhead{} &
        \multicolumn{4}{c}{Blue Filter (562~nm or 692~nm)} &  
        \multicolumn{4}{c}{Red Filter (832~nm or 880~nm)} \\
        \colhead{} & 
        \multicolumn{2}{c}{$\Delta m$} &
        \multicolumn{2}{c}{Mass ($M_{\rm J}$)} &
        \multicolumn{2}{c}{$\Delta m$} &
        \multicolumn{2}{c}{Mass ($M_{\rm J}$)} \\
        \colhead{Star} &
        \colhead{$0\farcs1$} &
        \colhead{$1\farcs0$} &
        \colhead{$0\farcs1$} &
        \colhead{$1\farcs0$} &
        \colhead{$0\farcs1$} &
        \colhead{$1\farcs0$} &
        \colhead{$0\farcs1$} &
        \colhead{$1\farcs0$} 
        }
    \startdata
    GL~317 & 4.66 & 6.93 & 159 & 159 & 4.45 & 5.96 & 159 & 159 \\
    HD~1326 & 2.59 & 6.56 & 202 & 167 & 3.56 & 8.07 & 167 & 167 \\
    HD~1388 & 3.02 & 5.56 & 721 & 546 & 3.96 & 6.74 & 557 & 184 \\
    HD~1461 & 2.12 & 5.39 & 808 & 517 & 3.77 & 6.99 & 472 & 186 \\
    HD~3765 & 4.41 & 7.14 & 420 & 203 & 3.58 & 6.10 & 203 & 203 \\
    HD~5319 & 3.37 & 6.02 & 872 & 612 & 3.26 & 7.03 & 669 & 554 \\
    HD~6558 & 2.82 & 5.00 & 864 & 616 & 3.50 & 8.31 & 633 & 270 \\
    HD~6734 & 2.79 & 6.05 & 937 & 553 & 3.79 & 7.89 & 593 & 456 \\
    HD~7924 & 2.56 & 4.46 & 644 & 481 & 3.36 & 6.20 & 311 & 178 \\
    HD~9986 & 3.51 & 6.62 & 653 & 353 & 3.27 & 7.54 & 574 & 181 \\
    HD~10436 & 2.67 & 4.91 & 554 & 279 & 3.31 & 6.14 & 201 & 201 \\
    HD~16160 & 3.11 & 6.63 & 555 & 192 & 3.90 & 6.63 & 192 & 192 \\
    HD~19522 & 3.18 & 4.98 & 847 & 662 & 3.90 & 6.70 & 604 & 336 \\
    HD~24040 & 4.06 & 5.52 & 633 & 558 & 3.98 & 5.83 & 531 & 196 \\
    HD~25311 & 4.28 & 6.50 & 723 & 569 & 3.93 & 7.15 & 659 & 248 \\
    HD~34445 & 2.53 & 4.15 & 819 & 647 & 4.18 & 6.06 & 504 & 209 \\
    HD~42618 & 2.72 & 4.50 & 735 & 555 & 3.64 & 6.50 & 482 & 192 \\
    HD~50499 & 4.95 & 8.66 & 548 & 218 & 4.61 & 8.22 & 302 & 218 \\
    HD~55696 & 5.02 & 8.66 & 573 & 253 & 4.44 & 8.67 & 509 & 253 \\
    HD~68017 & 3.62 & 6.57 & 620 & 306 & 3.00 & 7.40 & 580 & 167 \\
    HD~68988 & 3.60 & 5.51 & 683 & 550 & 3.79 & 6.84 & 578 & 188 \\
    HD~72490 & 3.57 & 5.32 & 887 & 728 & 3.54 & 6.02 & 775 & 525 \\
    HD~75732 & 3.21 & 6.58 & 613 & 212 & 5.19 & 7.15 & 212 & 212 \\
    HD~75898 & 3.62 & 6.11 & 753 & 568 & 3.52 & 6.84 & 669 & 246 \\
    HD~83443 & 5.51 & 8.60 & 237 & 237 & 4.97 & 7.08 & 237 & 237 \\
    HD~87359 & 2.52 & 5.26 & 719 & 519 & 4.54 & 6.57 & 228 & 186 \\
    HD~92788 & 2.55 & 5.69 & 776 & 434 & 3.93 & 6.62 & 421 & 216 \\
    HD~94834 & 5.52 & 7.96 & 658 & 553 & 3.56 & 7.43 & 641 & 553 \\
    HD~95735 & 3.83 & 6.91 & 162 & 162 & 2.75 & 6.24 & 162 & 162 \\
    HD~99491 & 3.40 & 5.78 & 640 & 406 & 4.01 & 6.50 & 281 & 202 \\
    HD~104304 & 5.43 & 7.89 & 434 & 216 & 3.65 & 8.07 & 390 & 216 \\
    HD~111031 & 1.24 & 5.86 & 942 & 471 & 3.27 & 7.32 & 615 & 234 \\
    HD~114174 & 3.14 & 4.99 & 675 & 545 & 4.00 & 5.58 & 492 & 184 \\
    HD~114783 & 2.55 & 4.18 & 650 & 495 & 3.99 & 6.09 & 181 & 181 \\
    HD~126614 & 2.28 & 5.31 & 774 & 527 & 1.93 & 6.25 & 774 & 265 \\
    HD~129814 & 3.57 & 5.58 & 672 & 469 & 3.45 & 6.16 & 531 & 234 \\
    HD~145675 & 2.92 & 5.11 & 659 & 447 & 3.53 & 7.43 & 398 & 222 \\
    HD~146233 & 2.88 & 4.31 & 716 & 578 & 2.92 & 6.04 & 609 & 181 \\
    HD~156668 & 3.33 & 6.84 & 556 & 193 & 2.79 & 6.14 & 372 & 193 \\
    HD~180053 & 3.81 & 5.98 & 882 & 642 & 4.30 & 6.49 & 579 & 579 \\
    HD~188015 & 4.14 & 7.73 & 593 & 207 & 3.36 & 6.76 & 531 & 207 \\
    HD~190406 & 3.24 & 5.43 & 702 & 534 & 3.47 & 6.63 & 590 & 176 \\
    HD~195564 & 2.51 & 5.99 & 874 & 520 & 3.33 & 7.68 & 656 & 250 \\
    HD~197076 & 2.88 & 5.39 & 698 & 533 & 3.45 & 6.74 & 499 & 198 \\
    HD~197162 & 3.28 & 6.49 & 963 & 578 & 3.18 & 6.96 & 728 & 578 \\
    HD~202696 & 3.14 & 5.39 & 1036 & 770 & 3.60 & 6.04 & 712 & 607 \\
    HD~207077 & 2.22 & 4.13 & 1181 & 826 & 2.68 & 5.07 & 869 & 551 \\
    HD~216520 & 1.83 & 3.48 & 704 & 570 & 3.54 & 5.58 & 283 & 176 \\
    HD~217850 & 3.22 & 5.03 & 691 & 585 & 3.19 & 5.67 & 615 & 271 \\
    HD~221354 & 3.30 & 5.34 & 601 & 400 & 3.56 & 6.71 & 322 & 199 \\
    HD~265866 & 2.62 & 4.79 & 147 & 147 & 3.75 & 6.05 & 147 & 147 \\
    HIP~52942~A & 2.78 & 5.84 & 884 & 596 & 3.56 & 7.23 & 725 & 242 \\
    HIP~57050 & 2.84 & 3.73 & 162 & 162 & 3.00 & 5.99 & 162 & 162 \\
    \enddata
    \tablenotetext{}{Notes. Magnitude limits are given in blue and red instrumental magnitudes. Reference Table \ref{tab:imaging} for the instrument used to observe each star.}
\end{deluxetable*}


\section{Literature Radial Velocity Solutions}

All of the stars in the sample have received some amount of RV characterization beginning with the analyses published by \citet{Butler2017}. Many of these stars have since been followed up through more detailed studies of the RV signals \citep[e.g.][]{Vogt2017,Ment2018,Luhn2019,Trifonov2019,Burt2021,Rosenthal2021}. Here, we focus on two categories of RV signals that are complemented by our speckle imaging data: unresolved, long-term RV trends and confirmed Keplerian signals from companions with minimum masses greater than 80~$M_{\rm J}$, roughly corresponding to the hydrogen burning limit. 

Of the 53 stars in our imaging sample, 4 have known companions with minimum mass greater than 80~$M_{\rm J}$ and 8 show unresolved trends in their RV time series that are not designated as stellar activity\footnote{\citet{Rosenthal2021} published linear trends for HD~34445 and HD~156668, which are members of our target sample. However, these trends are likely due to stellar activity as determined by emission in the Ca II H \& K lines measured in the same HIRES spectra as the RVs.}. These trends are quantified as a minimum RV semi-amplitude, which we list in Table \ref{tab:rv_trends}. The RV trends for seven of these eight stars were identified by the CPS Legacy Survey \citep{Rosenthal2021}. The eighth star with a trend, HD~19522, was not a member of the CPS Legacy Survey, so we measured the trend in the RV time series of HD~19522 as published by \citet{Butler2017}.   

\begin{deluxetable}{lc}
    \tablecaption{Summary of Systems with Unresolved RV Signals} 
    \tablecolumns{2}
    \tablewidth{0pt}
    \tablehead{  
        \colhead{Star} &
        \colhead{$\frac{1}{2}\Delta$RV (m~s$^{-1}$)} }
    \startdata
    HD~1388   & 230   \\
    HD~6734   & 172   \\
    HD~19522  & 657   \\
    HD~24040  &  24   \\
    HD~114174 & 655 \\
    HD~129814 & 297 \\
    HD~195564 & 377 \\
    HIP~57050 &   1 \\
    \enddata
    \tablenotetext{}{Notes. $\Delta$RV/2 is equal to half the total span of the RVs for a given target, which represents a minimum estimate of the RV semi-amplitude. For all stars except HD~19522, this value was derived from the RV acceleration parameters of \citet{Rosenthal2021}. For HD~19522, this value was directly measured from the RVs published by \citet{Butler2017}.}
    \label{tab:rv_trends}
\end{deluxetable} 

To constrain the objects causing the RV trends, we estimate the minimum RV semi-amplitude ($\Delta$RV/2) following \citet{Kane2014,Kane2019b}. $\Delta$RV/2 is simply half of the full range of RV observations. For HD~19522, we calculate this value directly. For the remaining nine stars, we use the published acceleration terms from \citet{Rosenthal2021} to reconstruct the RV trend before calculating $\Delta$RV/2. Table \ref{tab:rv_trends} lists $\Delta$RV/2 for each of the eight systems considered here.

From this lower limit on RV semi-amplitude, we calculate lower limits on companion mass ($M_c$) as a function of orbital semi-major axis ($a$) following
\begin{equation}\label{eq:rvsemiamp}
    \frac{\Delta \mathrm{RV}}{2} \le \sqrt{\frac{G}{a(1-e^2)}}
    \frac{M_c \sin{i}}{\sqrt{M_{\star}+M_c}}
\end{equation}

\noindent where $M_{\star}$ is the mass of the primary star and $G$ is the gravitational constant. In solving Equation \ref{eq:rvsemiamp}, eccentricity is drawn from a Beta distribution with shape parameters $s_{\alpha}=0.867$ and $s_{\beta} = 3.03$. This Beta distribution is motivated by empirical trends in the eccentricities of RV exoplanets \citep{Kipping2013a} and nearly indistinguishable from other empirically motivated distributions \citep[e.g., truncated Rayleigh;][]{Xie2016}. The value of orbital inclination we use in solving Equation \ref{eq:rvsemiamp} is drawn from a distribution uniform in $\cos{i}$. We take 5000 draws from these distributions, solving Equation \ref{eq:rvsemiamp} each time, to produce distributions of $M_c$ as a function of $a$. The black lines in Figures \ref{fig:imaging1} and \ref{fig:imaging1pt2} are the median values of these distributions, while the gray shaded regions illustrate the 16th and 84th percentiles. Since $\Delta$RV is a minimum semi-amplitude, the resulting $M_c$ values are minimum masses---but they are not $M\sin{i}$ values. The uncertainty in the $M_c$ values (i.e., the gray shaded regions) incorporate the unknown inclination and eccentricity of the companion's orbit.


\section{Results: Combining Imaging and RV Analyses}\label{sec:results}

The speckle imaging observations (Section~\ref{sec:speckle}) and the RV observations (Section~4) place upper and lower limits, respectively, on companion masses. Combining the two can potentially rule out substantial regions of parameter space in the mass--semimajor-axis plane that the companion could occupy. \citet{Montet2014} applied a rigorous likelihood analysis that involved marginalizing over eccentricity and inclination to determine the most probable masses and semi-major axes a companion could have, given the data. \citet{Kane2019b} applied a simplified variation on this approach by using the measured RV trend to place a limit on RV semi-amplitude. Here, we have followed the latter procedure as described in Section~4.

\begin{figure*}
  \begin{center}
    \begin{tabular}{cc}
      \includegraphics[width=8cm]{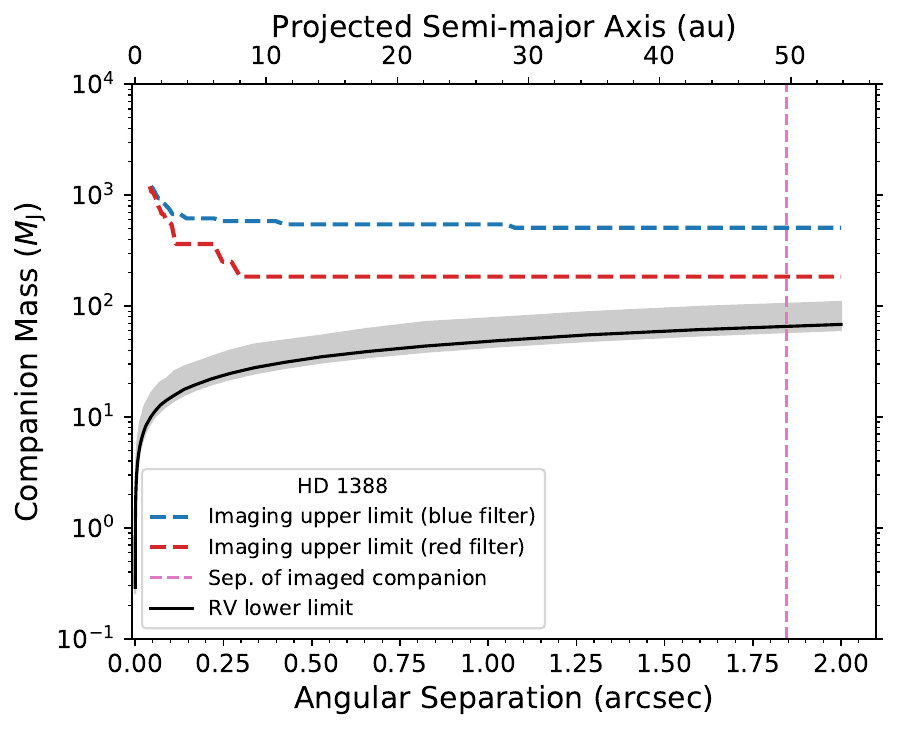} &
      \includegraphics[width=8cm]{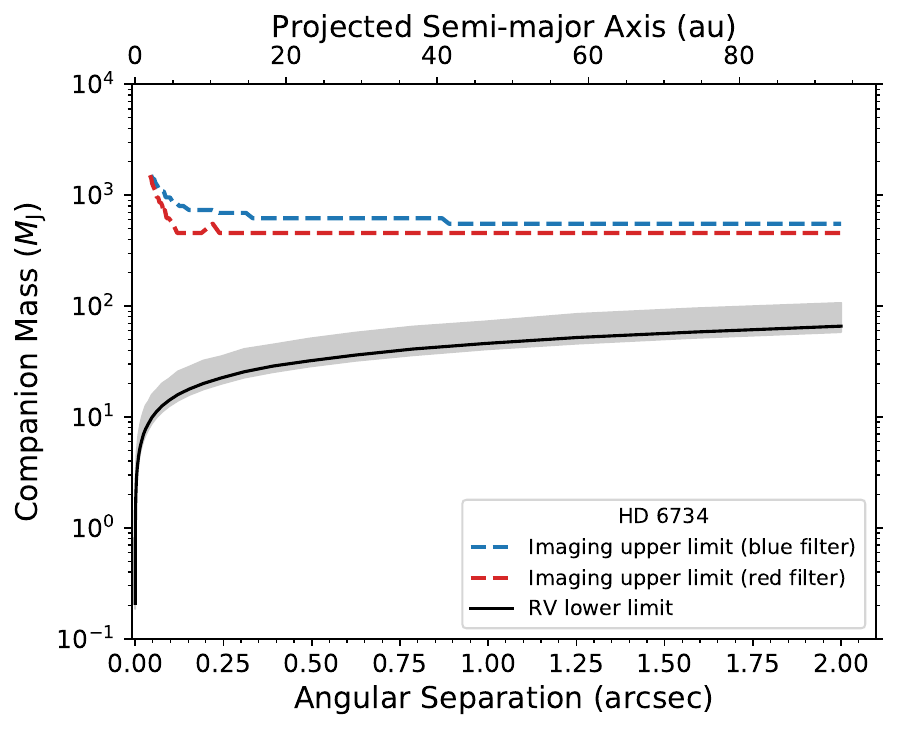} \\
      \includegraphics[width=8cm]{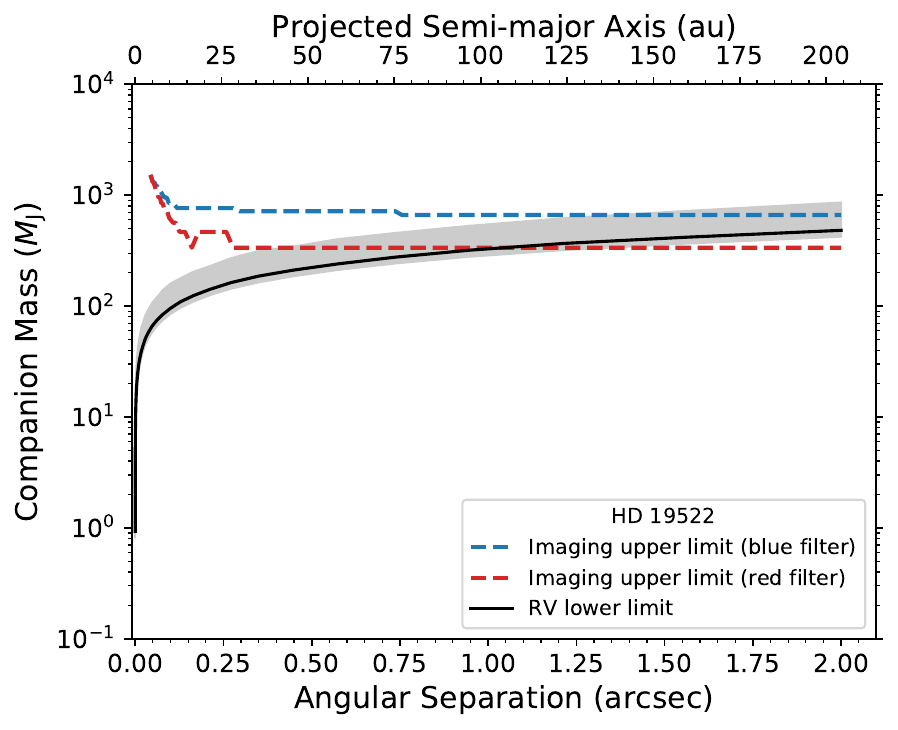} &
      \includegraphics[width=8cm]{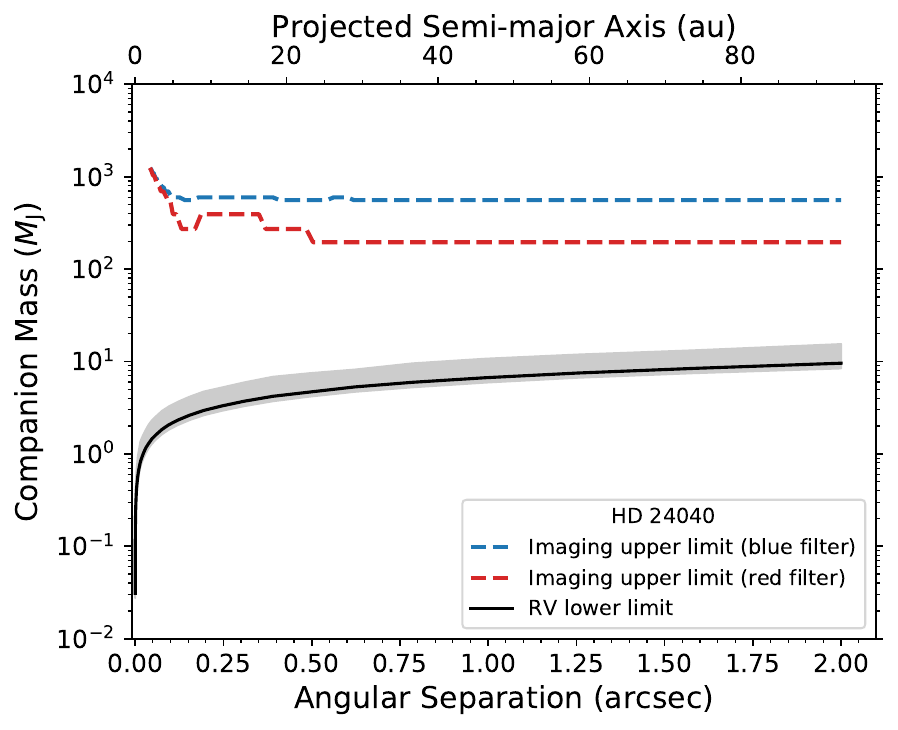} 
    \end{tabular}
  \end{center}
  \caption{Imaging and RV comparison for four of the eight systems with unresolved RV trends. The blue and red lines correspond to companion upper mass limits (ignoring any detected companion) based on the imaging observations. If a companion was detected, its separation is indicated by the vertical dashed line. The black line and gray shaded region correspond to the lower mass limit distributions (median and 16th--84th percentiles) for the object causing the RV trend. These distributions capture the uncertainty introduced by the unknown inclination and eccentricity of the object's orbit.}
  \label{fig:imaging1}
\end{figure*}

\begin{figure*}
  \begin{center}
    \begin{tabular}{cc}
      \includegraphics[width=8cm]{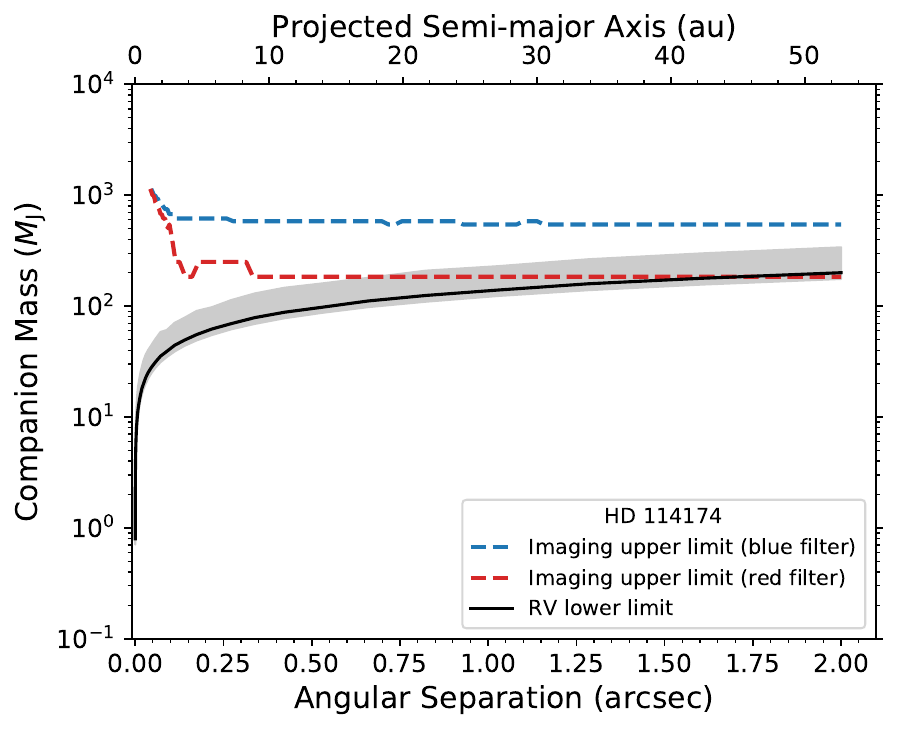} &
      \includegraphics[width=8cm]{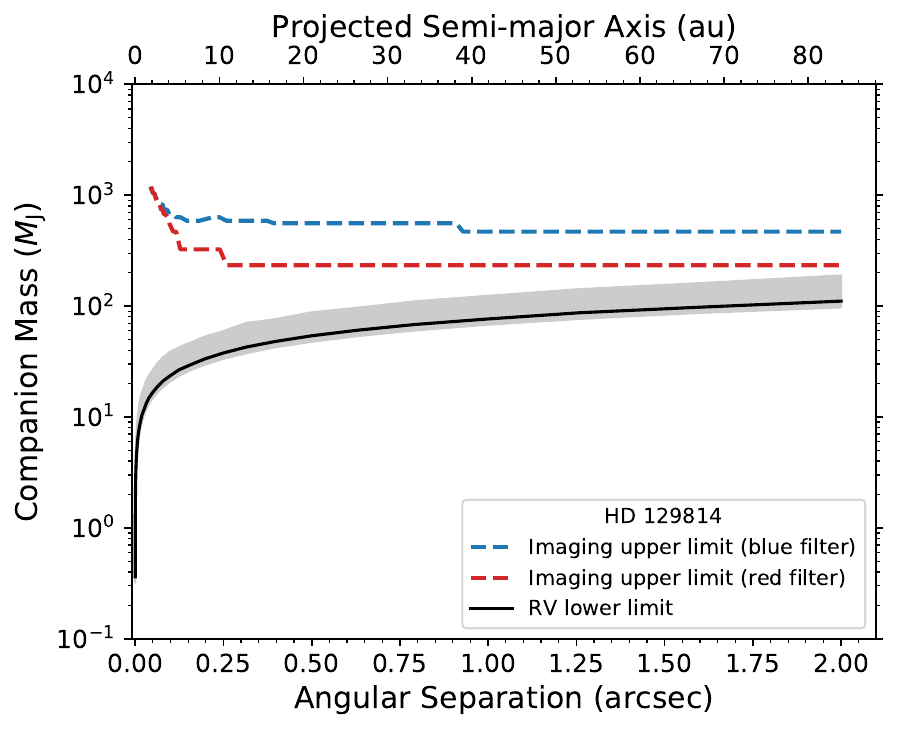} \\
      \includegraphics[width=8cm]{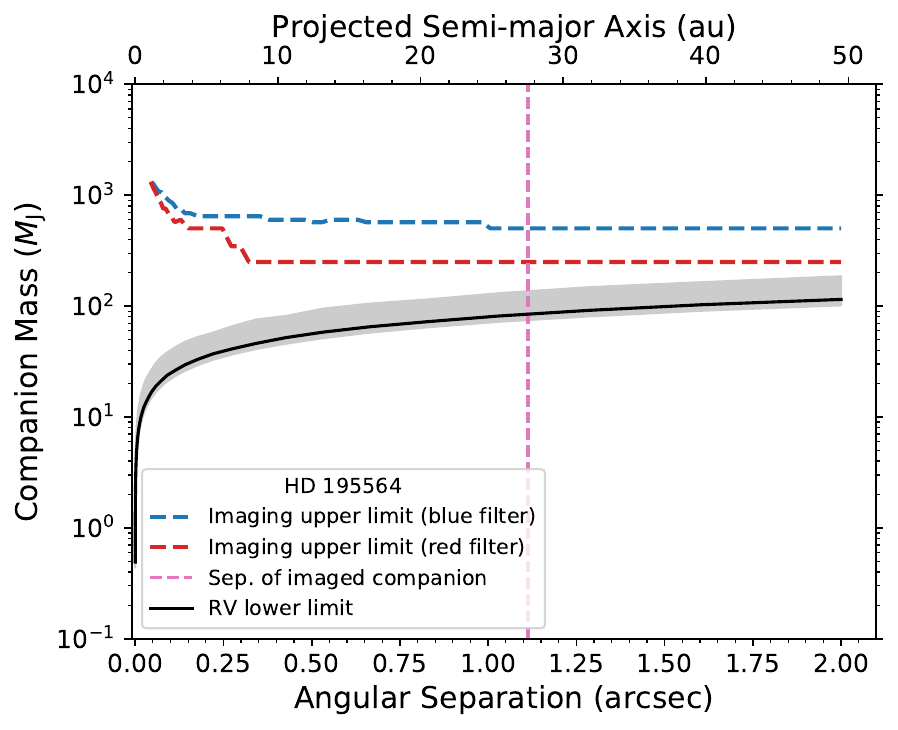} &
      \includegraphics[width=8cm]{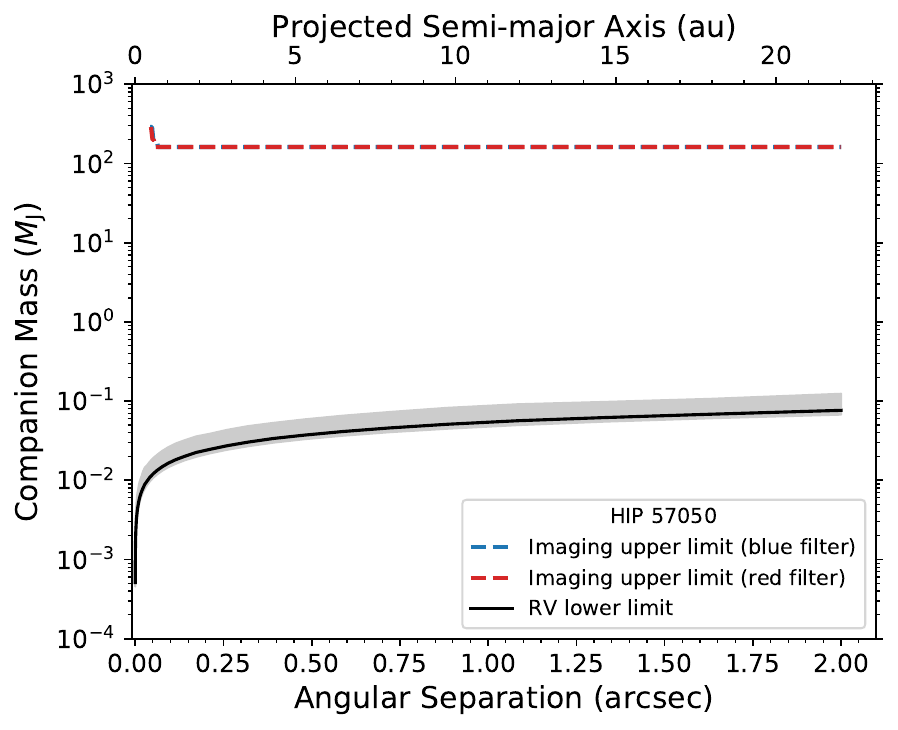} 
    \end{tabular}
  \end{center}
  \caption{Imaging and RV comparison for the remaining four systems with unresolved RV trends. The description is otherwise identical to Figure \ref{fig:imaging1}.\label{fig:imaging1pt2}}
\end{figure*}

\begin{figure*}
  \centering
    \begin{tabular}{cccc}
      \includegraphics[width=8cm]{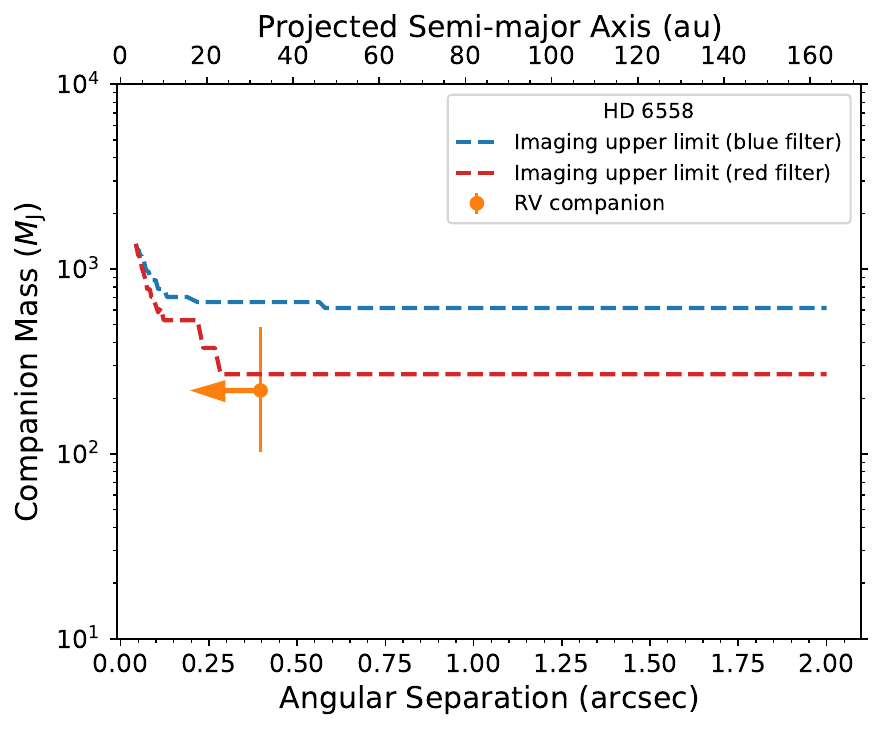} &
      \includegraphics[width=8cm]{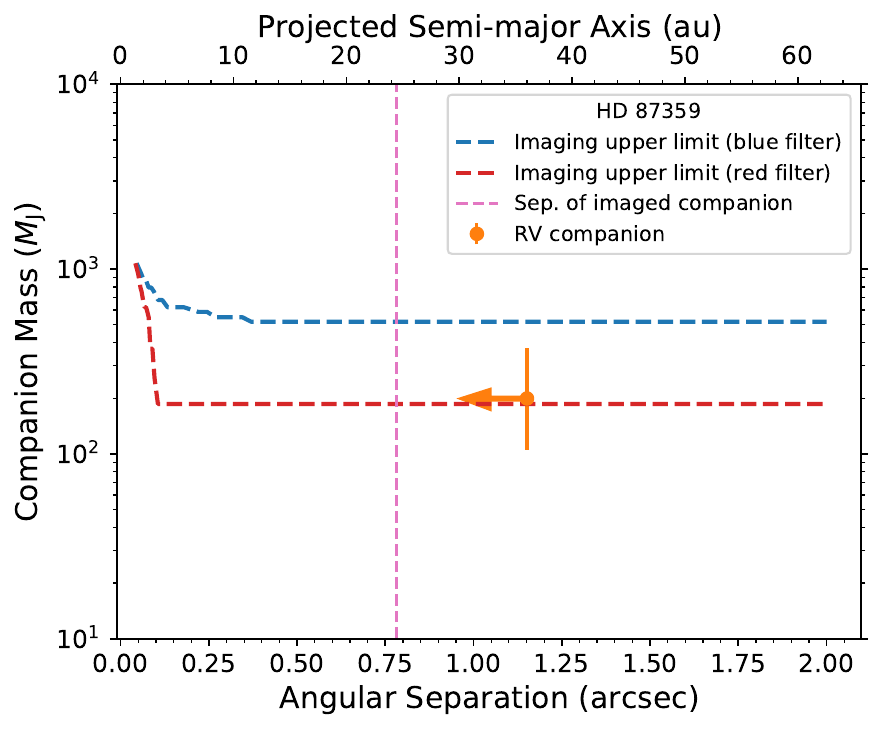} \\
      \includegraphics[width=8cm]{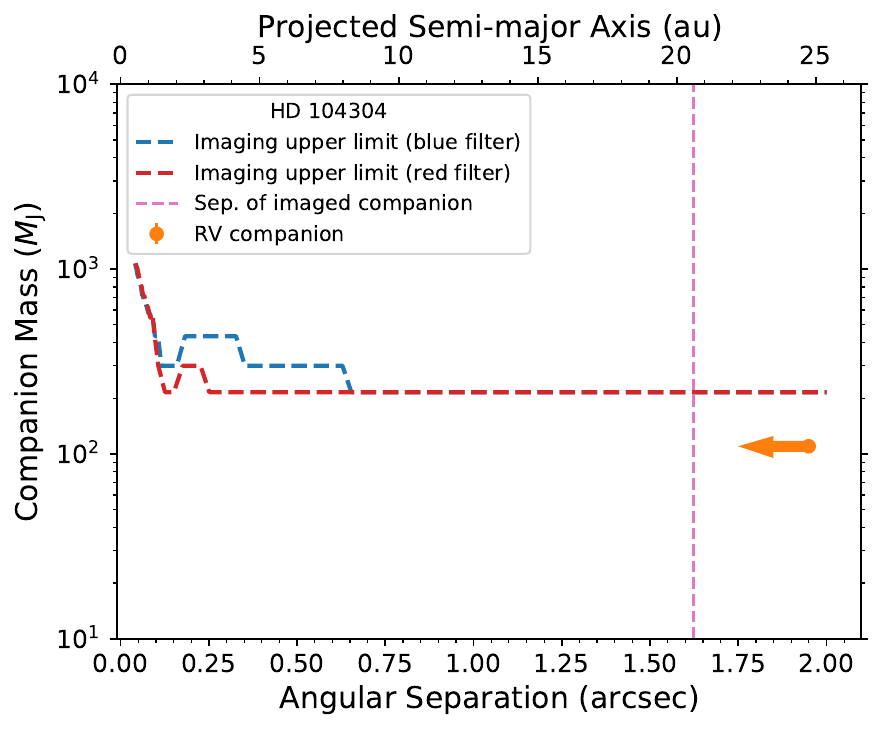} &
      \includegraphics[width=8cm]{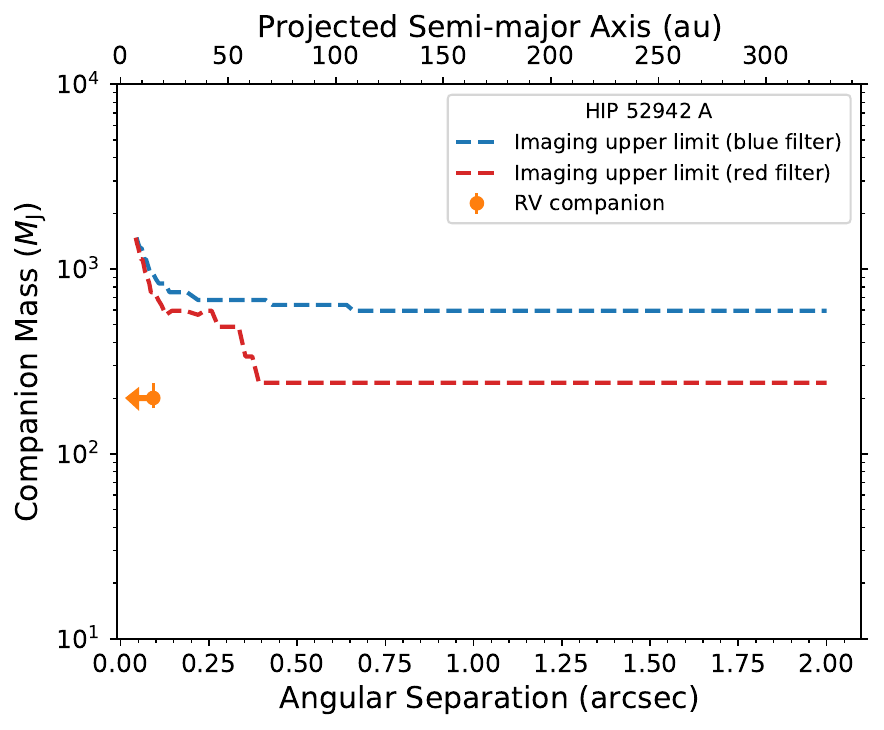} \\
    \end{tabular}
  \caption{Imaging and RV comparison for systems with RV companions with minimum mass greater than 80~$M_{\rm J}$. The blue and red lines correspond to companion upper mass limits (ignoring any detected companion) based on the imaging observations. If a companion was detected, its separation is indicated by the vertical dashed line. The orange data point shows the 68\% confidence interval for the mass of the RV companion. It is plotted at the maximum possible angular separation, given the known semi-major axis and orbital eccentricity. If a detected companion is within this separation, then it is consistent with the RV signal.}
  \label{fig:imaging2}
\end{figure*}

The lower limits on the companion mass from the RV data are calculated as functions of the semi-major axis ($a$), while the upper limits on the companion mass from the imaging data are calculated as a function of angular separation ($\alpha$). By placing both coordinates on the panels in Figures \ref{fig:imaging1}--\ref{fig:imaging2}, we make the assumption that these two coordinates are related by $\alpha = a/d$, where $\alpha$ is expressed in arcseconds, $a$ is expressed in astronomical units, and $d$ is the distance to the host star expressed in parsecs. This assumption neglects the unknown orbital inclination and longitude of periastron of any imaged companion. Therefore, special consideration is required in their interpretation. 

The combination of imaging and RV analysis for several systems with RV trends (Figures \ref{fig:imaging1} and \ref{fig:imaging1pt2}) places tight constraints on the properties of the companions causing the trends. Specifically, this applies to HD~19522, HD~114174, and HD~129814. For these systems, there is only a narrow area of mass--separation parameter space in which the companion can exist, and it is unlikely that the companion causing the RV trend is planetary or that it has an orbit with a maximum angular separations beyond $\sim 1 \arcsec$. 
Similarly, if the imaged companions we detected for HD~1388 and HD~195564 (Figures \ref{fig:imaging1} and \ref{fig:imaging1pt2}, vertical dashed lines) are responsible for the RV trends in these systems, then we find that their masses must be at least $\sim$60~$M_J$ and $\sim$90~$M_J$, respectively. This is consistent with our characterization of the detected companions (Table \ref{tab:comp_derived}).

For HD~6734, HD~24040, and HIP 57050, the combined imaging and RV constraints still leave a wide area of parameter space available for the properties of the companion. Improved characterization for this system and the others listed above through continued RV observations (i.e., until quadrature is observed) or deeper imaging observations would be useful. 

In the cases of the four systems with known RV companions of minimum mass 80~$M_J$ (Figure \ref{fig:imaging2}), we can make additional inferences about the companion properties. Since the orbital semi-major axes and eccentricities are known precisely, we can determine the maximum possible angular separations that the companions could have. In Figure \ref{fig:imaging2}, this is shown as an orange data point with an arrow pointing toward smaller separations. The error bars in the vertical axis are representative of the measured uncertainty in minimum companion mass. For HD~6558, the upper limits from the speckle imaging suggest that the 68\% confidence interval in minimum companion mass is representative of the likely companion's true mass (i.e., not just a lower limit). Therefore, the companion's orbit is more likely to be more edge-on ($\sin{i}\approx1$) than face-on ($\sin{i}\approx0$).

For HD~87359 and HD~104304, the RV companion minimum mass is known, and we detected a luminous companion (Figure \ref{fig:imaging2}, vertical dashed line). In both cases, the angular separation of the imaged companion is smaller than the maximum possible separation of the RV companion, suggesting that these two signals may indeed be caused by the same object. For both systems for which this is true, the companions have stellar masses.


\section{Discussion}\label{sec:discussion}

The transit method, while a successful avenue of planet discovery \citep[e.g.,][]{Thompson2018}, is severely limited by observational biases \citep[e.g.,][]{Beatty2008} such that only a few transiting planets on astronomical unit-scale orbits are known \citep[e.g.,][]{Wang2015,Kawahara2019,Dalba2019c}. RV surveys, on the other hand, maintain sensitivity out to wider orbital separation. Combined with decade-long baselines of stable observations, the longest running RV surveys \citep[e.g.,][]{Tinney2001,Butler2017,Wittenmyer2020,Rosenthal2021} are beginning to achieve sensitivity to signals resembling the giant planets in our solar system. Imaging follow-up to long-term RV surveys, such as we have presented here, plays a crucial role in validating the signals of wide-orbit planetary or stellar companions, especially if full orbits have not been resolved. By discovering previously unknown stellar companions and placing detection limits (Table \ref{tab:det_limits}) in systems with substantial RV acceleration, we have provided important information to supplement our understanding of planet occurrence on wide orbits. 

\begin{figure}
    \centering
    \includegraphics[width=\columnwidth]{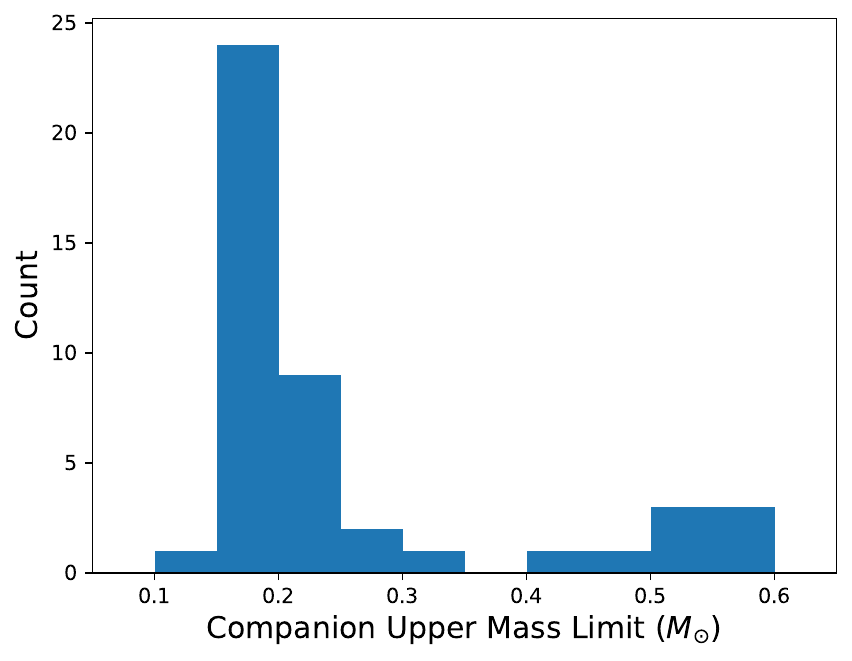}
    \caption{Upper mass limits for secondary stars at wide separations for systems without a detected companion. In 25 systems, we rule out secondaries with mass greater than 0.2~$M_{\sun}$.}
    \label{fig:upper_mass}
\end{figure}

Our results are broadly in line with similar efforts to combine imaging and RV data sets for the characterization of exoplanets,  \cite[e.g.,][]{Kane2014,Wittrock2016,Wittrock2017,Kane2019b}, brown dwarfs \citep[e.g.,][]{Crepp2016}, and white dwarfs \citep[e.g.,][]{Crepp2018,Kane2019a}. In our case, we have identified systems where the suspected planetary signal is fully consistent with a stellar companion (e.g., HD~1388, HD~87359, HD~104304) or where no companion is detected and the majority of the remaining parameter space for a companion is substellar and planetary (e.g., HD~24040, HIP~57050). For stars without detected companions at wide separations, the upper limit provided by the speckle imaging flattens to a single value. We show a histogram of these upper limits in Figure \ref{fig:upper_mass}. For 25 systems, we rule out companions with masses greater than 0.2~$M_{\sun}$, leaving only late-M dwarfs or substellar objects to explain the RV signals. In the interest of planet and brown dwarf discovery, these 25 systems, which can be identified in Table \ref{tab:det_limits}, should be prioritized for continued RV monitoring. 

Stellar binaries are a source of confusion for transit and RV surveys alike. For the former, pixels that subtend many arcseconds (as in the case of \TESS) can hide stellar companions leading to the inference of erroneous stellar and planetary properties, including radius and density \citep[e.g.,][]{Furlan2017b,Furlan2020,Ziegler2020}. For the latter, the unknown orbital inclination of objects causing long-term accelerations can allow stellar companions to contaminate exoplanet catalogs if an edge-on geometry is assumed \citep[e.g.,][]{Kiefer2021}. Therefore, direct imaging serves as a practical false-positive checking practice as well. 

The combination of imaging and RV data is a useful technique for exploring the properties of massive companions in nearby star systems, as demonstrated in this work and by others \citep[e.g.,][]{Kane2019b}. Including astrometric observations from missions such as {\it Hipparcos} and \Gaia\ \citep[e.g.,][]{Brandt2019} provides even more leverage in the characterization of companions. This includes testing for the unfortunate scenario of a photometric nondetection of an otherwise detectable companion that was near conjunction with the primary at the time of observation. Several of the systems in which we detected a stellar companion do have significant astrometric signals in {\it Hipparcos} and \Gaia\ data \citep{Brandt2018}. We leave the joint analysis of the RVs, imaging, and astrometry to a future analysis that will measure precise masses and orbits for these companions.


\section{Conclusions}\label{sec:conclusions}

We conducted speckle imaging of 53 stars using instruments at the WIYN and Gemini-South telescopes. The systems had previously been the targets of long-term RV campaigns \citep{Butler2017,Rosenthal2021}. We focused on systems that exhibited long-term RV trends or companions with measured minimum mass greater than 80~$M_J$, roughly corresponding to the hydrogen burning limit. However, we also provide companion detection limits in multiple filters for all of our imaging targets (Table \ref{tab:det_limits}). Our analysis yielded the following findings.
\begin{enumerate}
\item We detected luminous companions in our speckle images of eight systems including HD~1388, HD~25311, HD~87359, HD~104304, HD~111031, HD~126614, HD~146233, and HD~195564. HD~126614 \citep{Howard2010} and HD~195564 \citep[e.g.,][]{Burnham1879,Lloyd2002,Tokovinin2016} were known binary systems previously.
\item The color information we obtained for HD~146233 and HD~195564 enabled an isochrone analysis suggesting that the imaged companion in HD~195564 (at $\alpha=1\farcs1$) is likely gravitationally bound to the primary while the imaged companion in HD~146233 (at $\alpha=0\farcs$13) is likely not. Interestingly, neither the HST Guide Star catalog nor the USNO-B catalog detect a background star at the expected brightness or position that could explain the HD~146233 detection. We emphasize that it remains possible that our tentative detection for HD~146233 could be spurious. Further observations are needed to confirm this result.
\item In the cases of HD~1388, HD~87359, and HD~104304, the properties of the imaged companions are consistent with the RV measurements, providing support to the idea that these companions are associated with the primary stars.
\item For HD~6558, our speckle imaging observations provide evidence that the RV companion's orbital inclination is likely to be more edge-on rather than face-on.
\item In several systems with long-term RV trends (HD~19522 and HD~129814), we do not detect luminous companions through our speckle imaging, but the corresponding limits on companion mass rule out planetary scenarios.
\end{enumerate}
Our findings demonstrate the utility of synthesizing imaging and RV data sets for characterizing exoplanetary systems. For stars that are subject to long-term RV monitoring, the presence (or lack thereof) of stellar companions is vital information that is necessary to piece together the formation history of the system. The explanations of the RV trends we provide here are useful for understanding the population of exoplanets at wide orbits as well as making choices about future long-term RV monitoring efforts. Additionally, in all cases, the predictions made by our observations can be tested through additional RV monitoring and deeper imaging campaigns. 

During the preparation of this manuscript, \citet{Gonzales2020} identified neighboring stars in the HD~1388 and HD~111031 systems.


\acknowledgements
The authors thank the anonymous referee for the thoughtful comments that improved the quality and clarity of this work. P.D. acknowledges support from a National Science Foundation Astronomy and Astrophysics Postdoctoral Fellowship under award AST-1903811. G.W.H. acknowledges long-term support from NASA, NSF, Tennessee State University, and the State of Tennessee through its Centers of Excellence program. A portion of this research was carried out at the Jet Propulsion Laboratory, California Institute of Technology, under a contract with the National Aeronautics and Space Administration (80NM0018D0004). 

Based on observations obtained at the international Gemini Observatory, a program of NSF's NOIRLab, which is managed by the Association of Universities for Research in Astronomy (AURA) under a cooperative agreement with the National Science Foundation on behalf of the Gemini Observatory partnership: the National Science Foundation (United States), National Research Council (Canada), Agencia Nacional de Investigaci\'{o}n y Desarrollo (Chile), Ministerio de Ciencia, Tecnolog\'{i}a e Innovaci\'{o}n (Argentina), Minist\'{e}rio da Ci\^{e}ncia, Tecnologia, Inova\c{c}\~{o}es e Comunica\c{c}\~{o}es (Brazil), and Korea Astronomy and Space Science Institute (Republic of Korea). This research has made use of the NASA Exoplanet Archive, which is operated by the California Institute of Technology, under contract with the National Aeronautics and Space Administration under the Exoplanet Exploration Program. Some of the observations in the paper made use of the NN-EXPLORE Exoplanet and Stellar Speckle Imager (NESSI). NESSI was funded by the NASA Exoplanet Exploration Program and the NASA Ames Research Center. NESSI was built at the Ames Research Center by Steve B. Howell, Nic Scott, Elliott P. Horch, and Emmett Quigley.

\facility{Keck:I (HIRES), WIYN (NESSI), Gemini:South (DSSI)} 

\end{document}